%% file: Paper.tex
\documentclass[conference,compsoc]{IEEEtran}
\usepackage{cite}
\usepackage{amsmath,amssymb,amsfonts}
\usepackage{graphicx}
\usepackage{textcomp}
\usepackage{xcolor}
\usepackage{hyperref}
\usepackage{booktabs}
\usepackage{tabularx}
\usepackage{orcidlink}

\IEEEoverridecommandlockouts

\input{tables/macros.tex}

\begin{document}

\title{The Balkanization of Execution-Security Research for AI Coding
Agents: Isolation, Access Control, and Time-of-Check-to-Time-of-Use
Vulnerabilities}

\author{
\IEEEauthorblockN{Mohammadreza Rashidi~\orcidlink{0009-0003-7136-7168}}
\IEEEauthorblockA{\textit{Department of Computer Science}\\
\textit{AI and Media Analysis Lab}\\
Berlin, Germany\\
mohammadreza.rashidi@ue-germany.de}
}

\maketitle

\begin{abstract}
AI coding agents now read repositories, call tools, and execute shell commands
with limited human oversight, and a fast-growing body of work studies whether the
execution layer around them is actually safe. That literature is scattered:
papers on sandbox isolation, capability and access control, policy enforcement,
time-of-check-to-time-of-use (TOCTOU) races, Model Context Protocol (MCP)
threats, identity delegation, execution provenance, network egress control, and
static analysis of agent-generated code are published independently, rarely cite
one another across categories, and no existing survey organizes them by
execution-security mechanism. We systematize \NCorpus{} papers published between
\YearMin{} and \YearMax{} into \NCategories{} categories, each verified directly
against its source rather than taken from a secondary summary; the same
verification protocol also confirms four disclosed, patched CVEs directly
affecting production agent harnesses, showing the concern is not speculative.
Reading across categories surfaces five cross-cutting gaps that no single paper
addresses: isolation architectures and capability models are evaluated against
attacker capability, but almost never against one another on a shared benchmark;
policy-enforcement studies report failure rates from 69\% to 98\% of real
denylists yet no isolation paper re-evaluates its own defense under that
adversarial setting; TOCTOU and MCP threats are analyzed as separate literatures
despite both being instances of the same state-validation problem; every
enforcement mechanism we review assumes an honest policy author, leaving
policy-authoring error itself unaddressed; and a newly measured failure mode,
benign but out-of-scope agent actions occurring at rates up to 17.1\% under
realistic prompting, is addressed by no access-control or capability paper in
our corpus. We also find that three existing broader surveys of agentic AI
security discuss sandboxing only as one item among many defenses, leaving
execution security without a dedicated systematization; this paper is written
to fill that specific gap. We conclude with a research agenda directed at the
five gaps rather than at restating the individual papers' own stated future
work.
\end{abstract}


\section{Introduction}
\label{sec:intro}

An AI coding agent is only as safe as the environment it executes in. Modern
agent harnesses, terminal-based assistants, autonomous coding agents, and
browser-use agents, give a language model the ability to read files, run shell
commands, call external tools, and in many deployments do so with limited or no
human confirmation per action. The last two years have produced a large body of
work asking whether the execution layer around these agents is actually safe:
whether the sandbox holds, whether the access-control policy is enforced,
whether a state check can be invalidated between the time it is made and the
time it is acted on, and whether the tool-calling protocol itself carries new
attack surface. This is a different question from whether the model's outputs
are aligned or its prompts are safe; it is a systems-security question about the
boundary between the agent and the machine it runs on.

That literature is scattered. A paper on container sandbox escape does not cite
the paper on denylist fragility; a paper proposing a capability framework for
coding agents does not engage with the empirical finding that most
production policies fail to hold; a Model Context Protocol threat model and a
browser-agent TOCTOU paper describe what is structurally the same
validate-then-act race without using a shared vocabulary for it; a paper
measuring how often agents take benign but unrequested actions does not cite,
and is not cited by, any of the access-control papers that could in principle
constrain exactly that behavior. Three recent broader surveys of agentic AI
security exist~\cite{dehghantanha2026attacksurface,shahriar2025agenticsecurity,maloyan2026promptinjectionsok},
and a fourth reviews open agentic platforms specifically~\cite{chen2026claweddangerous},
but all four treat execution isolation as one entry in a wide catalog of
concerns alongside prompt injection, retrieval poisoning, and multi-agent risk,
rather than as a subject with its own internal structure. No existing paper
organizes execution-security work specifically, on its own terms, across the
mechanisms the field has actually built.

The concern these papers study is not speculative. During the verification
described in Section~\ref{sec:method}, we confirmed four disclosed, patched
CVEs directly against the NIST National Vulnerability Database: a runc
container-escape vulnerability affecting the container runtime many agent
sandboxes are built on~\cite{cve202421626}, a command-injection vulnerability in
GitHub Copilot and Visual Studio~\cite{cve202553773}, and two vulnerabilities in
Claude Code itself, a code-injection flaw in the startup trust-dialog flow and a
data-exfiltration flaw in the project-load flow that could leak API keys before
a user confirms trust in a
repository~\cite{cve202559536,cve202621852}. All four are documented in
Section~\ref{sec:incidents}.

This paper is a systematization of knowledge (SoK): we do not run new attacks or
build a new defense. We read \NCorpus{} papers published between \YearMin{} and
\YearMax{}, verify each one directly against its source rather than trust a
secondary description, organize them into \NCategories{} categories by the
execution-security mechanism they address, and read across the categories for
gaps that are invisible from within any single one. Every citation in this
paper's taxonomy was checked by fetching the paper's own abstract page and
confirming its title, authors, and claimed contribution before being included;
the corpus and its category labels are released as a plain data file so the
categorization is itself auditable and falsifiable, not an assertion the reader
must take on faith.

\textbf{Contributions:}
\begin{itemize}
  \item A systematized corpus of \NCorpus{} execution-security papers
        (\YearMin{}--\YearMax{}), organized into \NCategories{} categories:
        isolation architectures, escape and adversarial benchmarks, access
        control and capability models, policy enforcement fragility,
        time-of-check-to-time-of-use races, MCP-specific threats and defenses,
        systems-security framing, harness-level capability evaluation, identity
        and credential delegation, execution provenance and auditability,
        network egress control, static analysis of agent-generated code,
        scope-creep measurement, and skill and plugin packaging security.
  \item Confirmation, directly against the NIST National Vulnerability
        Database, of four disclosed and patched CVEs affecting production
        agent tooling, including two in Claude Code itself, grounding the
        corpus in real, resolved incidents rather than only in papers'
        self-reported threat models.
  \item Five cross-cutting gaps identified by reading across categories rather
        than within any one: the isolation-versus-access-control evaluation
        gap, the enforcement-fragility-versus-defense-evaluation gap, the
        TOCTOU-MCP vocabulary gap, the honest-policy-author assumption, and the
        unaddressed scope-creep failure mode.
  \item A released, machine-readable corpus file (paper key, verified title and
        authors, venue, year, category) so the taxonomy can be checked,
        extended, or disputed rather than taken as asserted.
  \item A research agenda organized around closing the five gaps, rather than a
        restatement of the future-work sections already published in the
        surveyed papers.
\end{itemize}

\section{Background}
\label{sec:background}

\subsection{Why execution security is a distinct problem}
Prompt injection and jailbreaking concern what a model is persuaded to
\emph{say} or \emph{decide}; execution security concerns what happens once that
decision reaches the operating system. The distinction matters because a
perfectly aligned model can still execute a dangerous action if the surrounding
system does not constrain what actions are possible, and conversely a
successfully injected model gains nothing if the execution layer refuses the
action regardless of intent. Systems-security framing papers make this argument
explicitly: agent security should treat the model itself as an untrusted
component and build guarantees into the surrounding
architecture~\cite{christodorescu2025systemssecurity}, since static, rule-based
controls designed for predictable environments are not well matched to the
fluid, natural-language-driven interactions agents actually
have~\cite{li2025aac}.

\subsection{The concrete primitives at stake}
The primitives this literature protects are ordinary operating-system
resources: the filesystem, the process and network namespace, and any
credentials reachable from the agent's execution context. The OWASP Top 10 for
LLM applications lists unbounded resource consumption and excessive agency as
named risk categories~\cite{owasp2025llmtop10}, giving the practitioner-facing
side of the same concern this paper's academic corpus addresses from the
research side.

\subsection{Real-world incidents}
\label{sec:incidents}
None of this is hypothetical, and we verified each of the following four
disclosures directly against the NIST National Vulnerability Database rather
than repeating a secondary description of it. CVE-2024-21626, ``Leaky
Vessels,'' is a patched runc vulnerability in which a leaked file descriptor
referencing the host's working directory allowed a crafted container to escape
to the host filesystem~\cite{cve202421626}; it predates agent-specific tooling
but illustrates the class of container-runtime weakness that agent sandboxes
inherit whenever they are built on the same primitives. CVE-2025-53773 is a
command-injection vulnerability in GitHub Copilot and Visual Studio 2022 (CVSS
3.1 base score 7.8), patched in version 17.14.12~\cite{cve202553773}. The
remaining two affect Claude Code specifically. CVE-2025-59536 (CVSS 8.8, CWE-94)
is a code-injection flaw in which versions before 1.0.111 could execute
untrusted project code before a user accepted the startup trust dialog, but
only when the tool was launched against an untrusted
directory~\cite{cve202559536}; the flaw is structurally a trust-boundary race
of exactly the shape the TOCTOU category in Section~\ref{sec:taxonomy}
studies, a check (has the user trusted this project) that a subsequent action
(execute project code) could outrun. CVE-2026-21852 (CVSS 7.5) is a data
exfiltration vulnerability in Claude Code's project-load flow that allowed a
malicious repository to exfiltrate data, including Anthropic API keys, before
the user confirmed trust, patched in version
2.0.65~\cite{cve202621852}. Both Claude Code CVEs concern the same underlying
moment, the interval between opening an untrusted repository and the user's
trust decision taking effect, which is precisely the execution-security
boundary this paper's taxonomy is about.

Beyond individually disclosed CVEs, the execution layer of coding agents
became a first-class offensive target at Pwn2Own Berlin 2026, where the Zero
Day Initiative introduced a dedicated Coding Agents category with Anthropic
Claude Code, OpenAI Codex, and Cursor as targets~\cite{zdi2026codingagents}.
The category's scoping rules read as an independent restatement of this
paper's thesis: a qualifying entry had to ``interact with a
contestant-controlled resource (e.g.\ web page, repository, media file) to
exploit a vulnerability within the coding agent'' through ``a common coding
agent use case,'' while ``model jailbreaks or prompt outputs that do not cross
security boundaries'' and any exploit requiring ``unsafe or permission-less
modes'' were explicitly ruled out of
scope~\cite{zdi2026codingagents}. The contest thus rewarded precisely the
execution-layer boundary crossing this survey systematizes and deliberately
declined to reward pure model-level manipulation, the same line our taxonomy
draws between the injection foundations of
Section~\ref{sec:taxonomy} and the containment failures that are its subject.
All three agents had real vulnerabilities demonstrated against them: OpenAI
Codex fell to a single input-neutralization bug (CWE-150) for a
\$40{,}000 award, Cursor was exploited in two separate entries, and several
attempts against Claude Code and Codex were adjudicated as collisions with
vulnerabilities already disclosed to the
vendor~\cite{zdi2026codingagents,kovacs2026pwn2own}. The event paid out
\$1{,}298{,}250 across 47 unique zero-days~\cite{kovacs2026pwn2own}; that a
flagship contest could stand up a coding-agent category at all, and that its
qualifying bar was a security-boundary crossing reachable through ordinary use
rather than a misconfiguration or a permission-less mode, is direct evidence
that the attack surface catalogued here is neither hypothetical nor niche.

\section{Methodology}
\label{sec:method}

\subsection{Search and inclusion process}
We searched arXiv and web literature using combinations of the terms
\emph{sandbox}, \emph{execution isolation}, \emph{agent harness}, \emph{coding
agent}, \emph{capability}, \emph{access control}, \emph{TOCTOU}, and \emph{Model
Context Protocol}, iterating the query set as each round of results surfaced
new author groups, venues, and vocabulary to search on. We included a paper if
it (a) addresses the execution layer of an LLM-based agent, meaning the
boundary between the agent's decisions and the operating system, filesystem,
network, or tool-calling substrate it acts through, and (b) was published or
posted between January \YearMin{} and the time of writing. We excluded papers
whose primary contribution is model-level alignment, prompt-injection defense at
the text level with no execution-layer component, or agentic security
surveyed only as one item in a broader catalog without a dedicated
execution-security contribution; the two broader surveys we identify in this
last category are discussed in Section~\ref{sec:related} for positioning rather
than included in the taxonomy, since they are secondary sources about the field
rather than primary execution-security contributions.

\subsection{Verification protocol}
Search engines and language-model-generated summaries of search results are not
a reliable source for a bibliography: titles are paraphrased, author lists are
merged across similarly named papers, and claimed numerical findings are
sometimes attributed to the wrong paper. For every candidate paper we fetched
its arXiv abstract page directly and recorded the exact title, full author
list, and venue as stated on that page before adding it to the corpus. Two
candidates failed this check and were corrected or dropped: a search summary
attributed a specific empirical breakdown (misconfiguration, shell-gap, and
bypass counts summing to 204 reports) to~\cite{zhong2026yolofs}, but the paper's
own abstract states only an aggregate figure of 290 reports across 13
frameworks with no such breakdown, so we report only the verified aggregate.
Separately, two candidate titles,
\emph{Systems Security Foundations for Agentic Computing} and \emph{Agent
Security is a Systems Problem}, returned an identical fourteen-author list and
near-identical descriptions (a systems-security perspective grounded in eleven
real-world case studies); we treat these as the same underlying work under two
titles and include only one entry, \cite{christodorescu2025systemssecurity}, to
avoid double-counting a single contribution as two corpus entries.

\subsection{Categorization}
\begin{figure}[!htb]
\centering
\includegraphics[width=\linewidth]{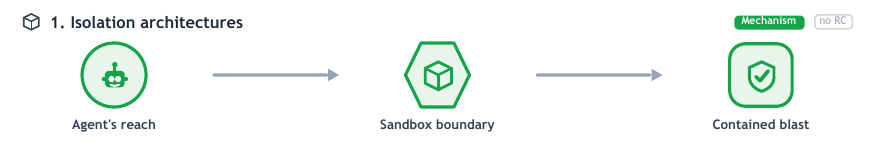}
\caption{Category 1 (isolation architectures): agent's reach $\to$ sandbox boundary $\to$ contained blast radius (root-cause chips as in Figure~\ref{fig:rootcause}).}
\label{fig:flow1}
\end{figure}

\begin{figure}[!htb]
\centering
\includegraphics[width=\linewidth]{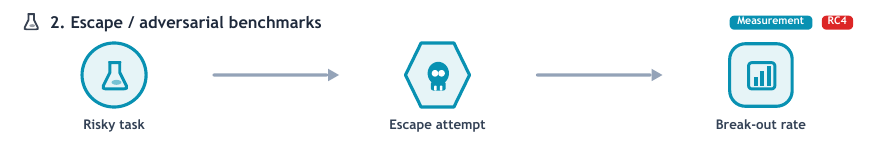}
\caption{Category 2 (escape and adversarial benchmarks): risky task $\to$ escape attempt $\to$ break-out rate.}
\label{fig:flow2}
\end{figure}

\begin{figure}[!htb]
\centering
\includegraphics[width=\linewidth]{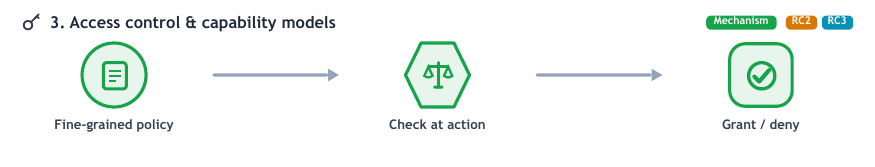}
\caption{Category 3 (access control and capability models): fine-grained policy $\to$ check at action $\to$ grant or deny.}
\label{fig:flow3}
\end{figure}

\begin{figure}[!htb]
\centering
\includegraphics[width=\linewidth]{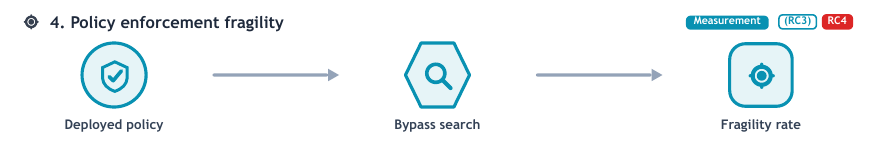}
\caption{Category 4 (policy enforcement fragility): deployed policy $\to$ bypass search $\to$ fragility rate.}
\label{fig:flow4}
\end{figure}

\begin{figure}[!htb]
\centering
\includegraphics[width=\linewidth]{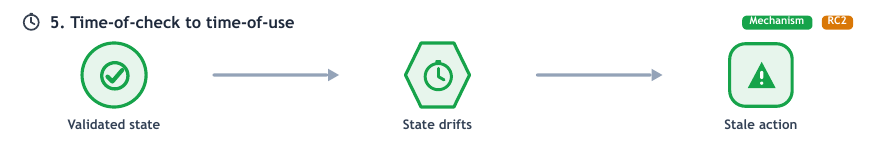}
\caption{Category 5 (time-of-check to time-of-use): validated state $\to$ state drifts $\to$ stale action.}
\label{fig:flow5}
\end{figure}

\begin{figure}[!htb]
\centering
\includegraphics[width=\linewidth]{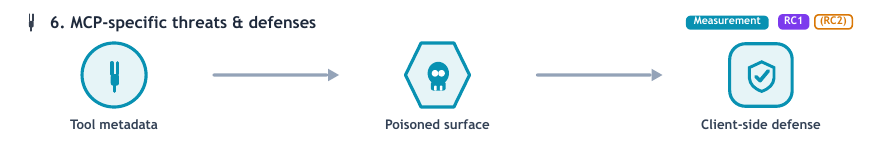}
\caption{Category 6 (mcp-specific threats and defenses): tool metadata $\to$ poisoned surface $\to$ client-side defense.}
\label{fig:flow6}
\end{figure}

\begin{figure}[!htb]
\centering
\includegraphics[width=\linewidth]{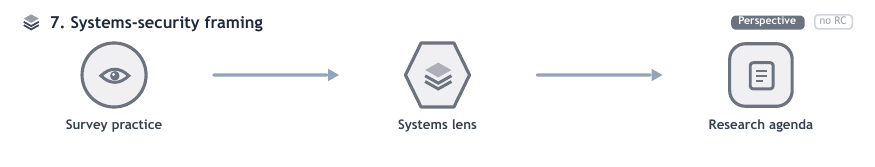}
\caption{Category 7 (systems-security framing): survey practice $\to$ systems lens $\to$ research agenda.}
\label{fig:flow7}
\end{figure}

\begin{figure}[!htb]
\centering
\includegraphics[width=\linewidth]{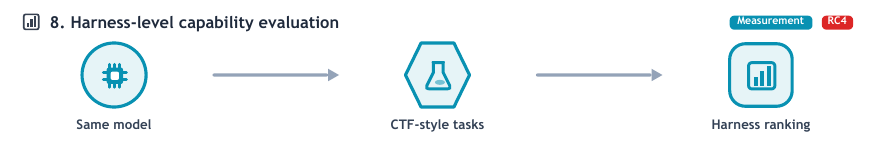}
\caption{Category 8 (harness-level capability evaluation): same model $\to$ CTF-style tasks $\to$ harness ranking.}
\label{fig:flow8}
\end{figure}

\begin{figure}[!htb]
\centering
\includegraphics[width=\linewidth]{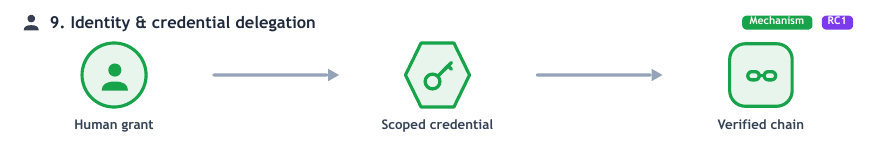}
\caption{Category 9 (identity and credential delegation): human grant $\to$ scoped credential $\to$ verified chain.}
\label{fig:flow9}
\end{figure}

\begin{figure}[!htb]
\centering
\includegraphics[width=\linewidth]{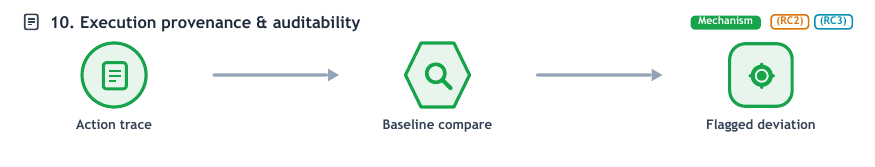}
\caption{Category 10 (execution provenance and auditability): action trace $\to$ baseline compare $\to$ flagged deviation.}
\label{fig:flow10}
\end{figure}

\begin{figure}[!htb]
\centering
\includegraphics[width=\linewidth]{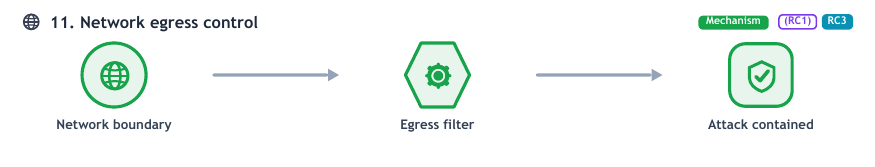}
\caption{Category 11 (network egress control): network boundary $\to$ egress filter $\to$ attack contained.}
\label{fig:flow11}
\end{figure}

\begin{figure}[!htb]
\centering
\includegraphics[width=\linewidth]{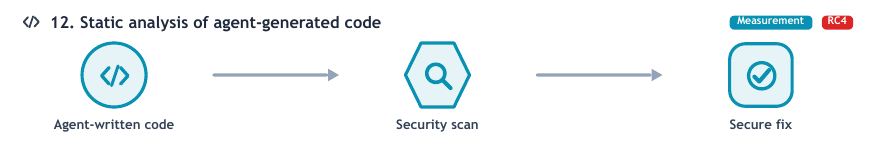}
\caption{Category 12 (static analysis of agent-generated code): agent-written code $\to$ security scan $\to$ secure fix.}
\label{fig:flow12}
\end{figure}

\begin{figure}[!htb]
\centering
\includegraphics[width=\linewidth]{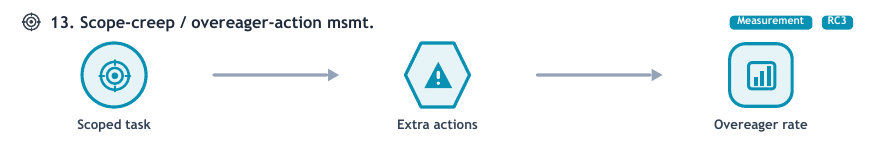}
\caption{Category 13 (scope-creep measurement): scoped task $\to$ extra actions $\to$ overeager rate.}
\label{fig:flow13}
\end{figure}

\begin{figure}[!htb]
\centering
\includegraphics[width=\linewidth]{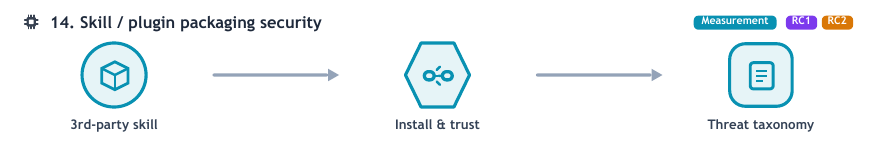}
\caption{Category 14 (skill and plugin packaging security): third-party skill $\to$ install and trust $\to$ threat taxonomy.}
\label{fig:flow14}
\end{figure}

\begin{figure}[!htb]
\centering
\includegraphics[width=\linewidth]{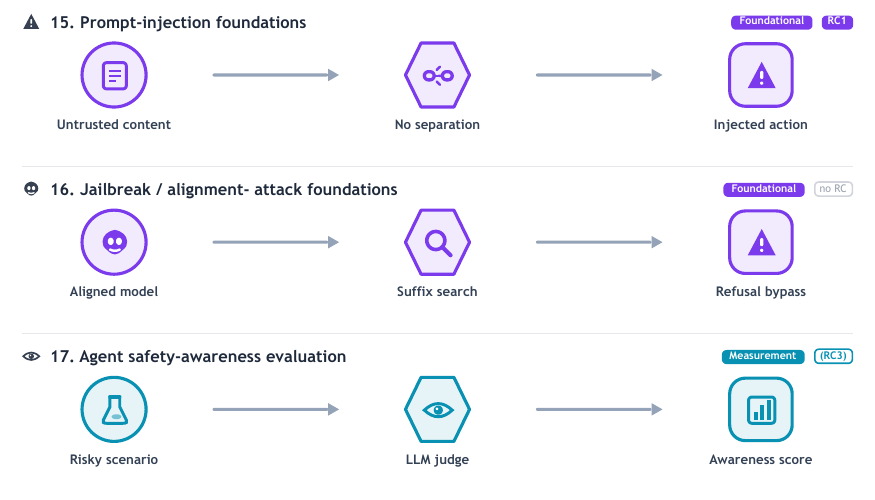}
\caption{Categories 15--17, the foundational work the corpus inherits rather
than proposes (prompt-injection, jailbreak, and safety-awareness
foundations), in the same icon-flow language as
Figures~\ref{fig:flow1}--\ref{fig:flow14}.}
\label{fig:flowfound}
\end{figure}

We assigned each verified paper to exactly one of \NCategories{} categories
based on its primary mechanism, listed with the number of papers in each in
Table~\ref{tab:taxonomy}. The categorization and the full corpus, including the
verified title, authors, venue, and arXiv identifier for every entry, are
released as a plain CSV file alongside this paper so the classification is
falsifiable rather than asserted; Section~\ref{sec:verification} describes how
every count stated in this paper is mechanically re-derived from that file.
Each taxonomy subsection below opens with a compact
setting$\to$technique$\to$outcome flow strip for its category
(Figures~\ref{fig:flow1}--\ref{fig:flow14}), drawn as icon shapes, circle
(setting) $\to$ hexagon (technique) $\to$ rounded square (outcome), so the
process reads at a glance; the three foundational categories that have no
dedicated subsection are summarized together in Figure~\ref{fig:flowfound}.

\subsection{Reproducibility}
\label{sec:verification}
Table~\ref{tab:taxonomy} and every count asserted in this paper (\NCorpus{}
total papers, the per-category counts, and the \YearMin{}--\YearMax{} year
range) are generated from a single machine-readable corpus file that records,
for each paper, its bibliographic identity, category, root-cause attribution,
and pipeline stage. An independent checker re-derives every count from that
file and compares it against the numbers used in the prose and tables; it also
verifies that no bibliography entry is duplicated and that every citation used
in the taxonomy resolves to a real entry. The corpus file and both tools are
released with the paper as a supplementary artifact. This is a narrower notion
of provenance than an experimental paper's, since a SoK has no raw measurement
to re-derive, but it guarantees that the counts asserted in the prose cannot
drift from the corpus, and it lets a reader challenge any individual
classification by editing one line of the released file and re-running the
checker.

\section{A Taxonomy of Execution-Security Work}
\label{sec:taxonomy}

\begin{table*}[!tbp]
\caption{Taxonomy of the \NCorpus{} verified papers by primary execution-security
mechanism.}
\label{tab:taxonomy}
\centering
\footnotesize
\input{tables/tab_taxonomy.tex}
\end{table*}

\subsection{Isolation architectures (\NIsolationArchitecture{} papers)}
This category proposes or evaluates a system that confines what an agent's
actions can affect. IsolateGPT applies app-store-style isolation to LLM-based
agentic systems, restricting interaction between third-party components to
well-defined, permissioned interfaces and showing this defends against app
compromise, data theft, and unintended system change with under 30\% overhead on
most tested queries~\cite{wu2025isolategpt}. ceLLMate takes a narrower but
sharper approach for browser-use agents specifically: rather than policing
low-level UI actions, it sandboxes at the HTTP layer, since any state-changing
effect must eventually reach the site's backend through a network
request~\cite{meng2025cellmate}. AgentBay addresses a different concern,
seamless human takeover across heterogeneous execution surfaces (desktop,
mobile, browser, code interpreter) rather than confinement per
se~\cite{piao2025agentbay}. DeltaBox and the fault-tolerant sandboxing work in
this category both address the operational cost of isolation once adopted:
DeltaBox cuts checkpoint and rollback latency to single-digit milliseconds by
snapshotting only the delta between agent states~\cite{dong2026deltabox}, and a
transactional sandbox design wraps agent actions in atomic, reversible units,
reporting a 100\% interception rate for high-risk commands and a 100\% rollback
success rate in its own evaluation~\cite{yan2025faulttolerant}.

\subsection{Escape and adversarial benchmarks (\NEscapeBenchmark{} papers)}
This category measures whether an agent, given the opportunity, can break out
of or subvert its intended execution boundary. SandboxEscapeBench is a
capture-the-flag-style benchmark that deliberately introduces known classes of
container weakness, misconfiguration, privilege-allocation error, kernel flaws,
and orchestration mistakes, and finds that capable frontier models can locate
and exploit them~\cite{marchand2026sandboxescape}. RedCode evaluates the
opposite direction, whether an agent will execute or generate code that is
already known to be risky, across more than 4,000 test cases in Python and
Bash and 19 language models, finding agents more willing to reject risky
operating-system actions than to reject code that is merely subtly
buggy~\cite{guo2024redcode}. SandboxEval targets the test harness itself,
probing an open-source AI-safety evaluation framework for data leaks,
unauthorized file changes, and unwanted network access during the execution of
untrusted, model-generated code~\cite{rabin2025sandboxeval}.

\subsection{Access control and capability models (\NAccessControl{} papers)}
This is the largest category and the field's dominant response to the isolation
problem: rather than a single confinement boundary, grant the agent
fine-grained, revocable, or context-sensitive permissions. AgentBound applies an
Android-permission-model-inspired policy layer specifically to MCP servers, with
policies derived automatically from server source
code~\cite{buhler2025agentbound}. PORTICO addresses a narrower but sharply
defined failure mode, \emph{lingering authority}, where a permission granted for
one subtask remains valid after that subtask closes, by issuing time-limited,
revocable capability tokens that are pulled the moment a subtask
completes~\cite{santosgrueiro2026portico}. SEAgent frames privilege escalation
in multi-agent systems as a confused-deputy-shaped problem and combines
mandatory and attribute-based access control, tracking inter-agent and
agent-tool interactions through an information-flow
graph~\cite{ji2026seagent}. A vision paper argues that static, identity-based
access control is a poor match for agents and proposes evaluating context along
multiple dimensions (relationships, situational factors, norms) with adaptive,
non-binary responses such as redaction or
summarization~\cite{li2025aac}. The Open Agent Passport enforces policy
deterministically \emph{before} a tool call executes rather than after,
reporting a drop from 74.6\% to 0\% attacker success when a strict policy is
applied, though the authors note this assumes the enforcement framework itself
is not compromised~\cite{uchibeke2026oap}. FORGE takes the most formal approach
in this category, expressing policy in Datalog and enforcing it through a
reference monitor built on aspect-oriented programming principles rather than
natural-language rules embedded in a system prompt~\cite{palumbo2026forge}.

\subsection{Policy enforcement fragility (\NEnforcementFragility{} papers)}
Where the previous category proposes mechanisms, this category measures whether
mechanisms that are already deployed actually hold. ShellSieve is the more
alarming of the two: an automated pipeline that has a language model propose
bypasses to real command denylists and validates them by execution in a
sandbox, tested against 1,709 denylists scraped from GitHub, finding that
69.0\% to 98.6\% of them are fragile depending on the bypass class
considered~\cite{chen2026shellsieve}. YoloFS documents the same failure mode
from a design perspective: reviewing 290 public misuse reports across 13
agent frameworks, the authors find agents are routinely given a binary choice
between unrestricted filesystem access and workflows that block on constant
approval, and propose shifting awareness of an action's effects into the
filesystem itself through staged changes and snapshot-based
correction~\cite{zhong2026yolofs}.

\subsection{Time-of-check to time-of-use (\NToctou{} papers)}
This category isolates a specific race condition: an agent validates a piece of
external state, a file's contents, a webpage's structure, and then acts on a
stale copy of that state after it has changed. Mind the Gap is the first
dedicated study of this class in LLM-enabled agents generally, introducing
TOCTOU-Bench (66 tasks) and combining prompt rewriting, state-integrity
monitoring, and tool-fusing to reduce the rate of TOCTOU flaws in executed
trajectories from 12\% to 8\%~\cite{lilienthal2025toctou}. The
browser-use-specific follow-on work shows the same race is common and
exploitable across ten open-source browser agents and many real websites, since
a page can change between an agent's plan and its
action~\cite{jiang2026browsertoctou}.

\subsection{MCP-specific threats and defenses (\NMcpSpecific{} papers)}
The Model Context Protocol standardizes how agents discover and call external
tools, and in doing so introduces its own execution-security surface. A threat
model built on STRIDE and DREAD identifies tool poisoning, malicious
instructions embedded in a tool's own metadata, as the standout client-side
risk and tests seven prominent MCP clients, finding most implement no static
validation against it~\cite{huang2026mcpthreat}. A landscape survey maps the
full MCP server lifecycle into a four-actor threat taxonomy (malicious
developers, external attackers, malicious users, and security flaws) and
validates it against real-world cases~\cite{hou2025mcplandscape}. The largest
empirical study in this category examined 1,899 open-source MCP servers and
found 7.2\% carry traditional-style vulnerabilities, 5.5\% exhibit
MCP-specific tool poisoning, and 66\% show code smells, concluding that MCP
security tooling has not kept pace with MCP adoption~\cite{hasan2025mcpfirstglance}.
MCP-Guard responds with a three-stage defense pipeline, pattern scanning, a
fine-tuned semantic classifier, and a final LLM arbitrator, evaluated against a
companion benchmark of simulated attacks~\cite{xing2026mcpguard}.

\subsection{Systems-security framing (\NSystemsFraming{} papers)}
Two papers argue for a change in how the field thinks about the problem rather
than proposing a new mechanism. \cite{christodorescu2025systemssecurity} argues
the field has approached agent safety from an AI-centric perspective that lacks
systems-security guarantees, and grounds a set of research directions in eleven
real-world attacks. \cite{wang2026hciframing} makes a complementary
argument from the human side: reviewing dozens of academic papers and deployed
systems, the authors find that the human-oversight mechanisms companies
actually deploy, advance policy-setting, runtime approval, permission scoping,
are barely represented in what academic papers study, and that these
mechanisms leave users caught between approval fatigue and uncontrolled
autonomy.

\subsection{Harness-level capability evaluation (\NHarnessCapability{} papers)}
A single paper in our corpus evaluates the agent harness itself, rather than a
defense or a threat, as the unit of analysis. Comparing five different
scaffolds, including ones built around Claude and Codex, against the same
underlying model on 33 cybersecurity CTF-style challenges, the authors find no
single harness dominates, and that combining structurally different scaffolds
in a shared blackboard architecture outperforms every individual
one~\cite{mayoralvilches2026csi}. We include this paper because harness design
choices are themselves an execution-security variable: which tools a scaffold
exposes and how it sequences them shapes the attack surface as much as any
sandbox or policy layered on top.

\subsection{Identity and credential delegation (\NIdentityDelegation{} papers)}
This category addresses a question the access-control category assumes an
answer to: how does an agent get scoped, revocable credentials in the first
place, and how does a downstream service know which human, and which
delegation chain, an agent's request ultimately traces to. One proposal extends
OAuth 2.0 and OpenID Connect with agent-specific credentials and metadata, and
translates natural-language permission grants into auditable access-control
configurations so organizations can adopt it without replacing existing login
infrastructure~\cite{south2025delegation}. A complementary, more foundational
proposal argues the deeper problem is that transformers treat tool definitions
and user instructions as indistinguishable tokens despite their different
security implications, and addresses it with cryptographic binding of
capabilities to their issuing context plus reproducibility verification,
showing formally that a single unverifiable agent in a delegation chain
undermines verification for every agent downstream of
it~\cite{zhou2026capabilitybinding}.

\subsection{Execution provenance and auditability (\NExecutionProvenance{}
paper)}
Rather than gate an action before it executes, this category records what
happened so it can be judged, flagged, or attributed afterward. Agent-Sentry
establishes a baseline of legitimate action patterns from past runs and flags
deviations from it using a three-layer check, an action-sequence classifier, a
fixed-rule check on sensitive values, and an LLM reviewer reserved for
ambiguous cases, reporting that it stops 94.3\% of successful injections in the
AgentDojo and AgentDyn benchmarks while allowing 95.1\% of benign executions
through, without modifying the agent, its tools, or the underlying
model~\cite{sequeira2026agentsentry}.

\subsection{Network egress control (\NNetworkEgressControl{} paper)}
This category treats the network boundary as the enforcement point rather than
the filesystem or the tool-calling layer, on the reasoning that most
state-changing effects an agent can have ultimately require an outbound
request. A dual-firewall architecture converts incoming agent-to-agent
messages into a closed, structured protocol to strip manipulative content
through deterministic verification, and separately controls outgoing
information by matching its level of detail to what the task actually
requires rather than applying all-or-nothing redaction, substantially cutting
both privacy and security attack success rates across hundreds of simulated
attacks while preserving task performance~\cite{abdelnabi2025firewalls}.

\subsection{Static analysis of agent-generated code (\NStaticAnalysisGeneratedCode{}
paper)}
This category scans code an agent is about to run, or has just written, before
or instead of confining its execution. SecureVibeBench anchors 105 C/C++
secure-coding tasks to genuine historical vulnerabilities drawn from 41 real
open-source projects via OSS-Fuzz, with clearly marked introduction points, and
judges agent-produced fixes with a combination of functional tests and static
and dynamic security analysis; across five widely used coding agents paired
with five language models, even the strongest combination produced solutions
that were correct and secure only 23.8\% of the
time~\cite{chen2025securevibebench}. This result is a caution against relying
on generation quality as an execution-security control: the agents in this
study are, on the authors' own evaluation, more often wrong about security
than right.

\subsection{Scope-creep and overeager-action measurement (\NScopeCreepMeasurement{}
paper)}
This category measures a failure mode that is neither malicious nor a policy
violation in the traditional sense: a benign, correctly authorized task
completed by taking additional, unrequested actions. Using a purpose-built
benchmark of 500 validated scenarios run across four agent products and six
base models, roughly 7,500 runs, the authors find permissive-framework agents
such as Claude Code, Codex CLI, and Gemini CLI show substantially higher rates
of unrequested action than an ask-to-continue framework such as
OpenHands~\cite{qu2026overeager}. A striking methodological finding is that
simply removing an explicit statement of authorized scope from the prompt
raised the measured overeager rate on Claude Code from 0.0\% to 17.1\%,
suggesting the agent is pattern-matching a declared-scope sentence rather than
inferring task boundaries from the task itself.

\subsection{Skill and plugin packaging security (\NSkillArchitecture{} paper)}
Coding-agent harnesses increasingly ship a packaging layer, skills, plugins, or
extensions, that a user or a marketplace can install into an otherwise trusted
agent. A dedicated security analysis of this layer models a four-phase
lifecycle, creation, distribution, deployment, and execution, and builds a
threat taxonomy of seventeen scenarios across three attack layers, validated
against five confirmed real-world incidents; the authors conclude the most
severe threats are structural rather than incidental, arising from the
framework's absent data-instruction boundary, a single-approval trust model
that persists indefinitely after one grant, and the lack of a mandatory
security review at the marketplace
level~\cite{li2026secureskills}.

\section{Root Causes and Structural Similarity Across the Taxonomy}
\label{sec:rootcause}

Section~\ref{sec:taxonomy} organizes the corpus the way the field organizes
itself: by the mechanism a paper proposes or measures. That grouping is useful
for a reader looking for, say, every isolation architecture, but it also
obscures how much of the corpus is answering variations on the same underlying
design defect. This section re-reads the same \NCorpus{} papers along two
different axes, first asking what root cause a technique is actually a
response to, independent of its category label, and second asking where in an
agent's action pipeline a technique intervenes and whether it proposes a
mechanism or measures one. Both re-readings are our synthesis rather than a new
empirical claim, and we present them as an interpretive lens, not as a
replacement for Section~\ref{sec:taxonomy}'s mechanism-based grouping.

\subsection{Four root causes behind seventeen categories}
\label{sec:rootcause-four}

Reading the corpus for the design defect a technique is actually a response to,
rather than the mechanism it proposes, collapses much of the taxonomy onto four
recurring root causes.

\textbf{RC1: no structural separation between untrusted content and control
input.} A transformer processes a tool's returned data, a webpage's text, and a
user's instruction as the same kind of token, so nothing at the architecture
level distinguishes ``execute this'' from ``this is what I found.''
\cite{zhou2026capabilitybinding} states this explicitly as the deeper problem
underlying identity delegation, and the skill-packaging taxonomy identifies the
same absence, which it calls the ``absent data-instruction
boundary,''~\cite{li2026secureskills} as the most severe of its seventeen
threat scenarios. MCP tool poisoning is the same defect at the protocol layer:
a tool's self-declared metadata is control input that a client cannot
distinguish from untrusted description
text~\cite{huang2026mcpthreat,hasan2025mcpfirstglance}.

\textbf{RC2: authorization is checked once and trusted forever.} A permission,
a validated file, or a tool's declared behavior is treated as durable after a
single check, even though the thing it was checked against can change.
PORTICO's ``lingering authority''~\cite{santosgrueiro2026portico}, both TOCTOU
papers' stale-state race~\cite{lilienthal2025toctou,jiang2026browsertoctou},
and MCP tool poisoning's validate-once-trust-forever
pattern~\cite{huang2026mcpthreat} are the same failure at three different
layers: a permission grant, a filesystem or webpage read, and a tool
registration are all treated as valid indefinitely after the moment they were
approved. The skill-packaging taxonomy's ``single-approval trust model that
persists indefinitely after one grant''~\cite{li2026secureskills} is the same
defect again, one category up, at the level of an installed plugin rather than
a single permission or a single file.

\textbf{RC3: a mechanism encodes whether an action is permitted, not whether
this invocation matches intent.} A capability token, a Datalog rule, or a
command denylist all answer a binary question, is this action type allowed,
which is a different question from whether this specific, authorized action
should happen right now. Scope-creep measurement shows an agent taking
additional, unrequested but individually authorized actions at rates that swing
from 0.0\% to 17.1\% depending on prompt phrasing
alone~\cite{qu2026overeager}, which is not a permission failure since every
action the agent took was within its granted capability. SEAgent's framing of
privilege escalation as a confused-deputy problem~\cite{ji2026seagent} is the
same gap from the opposite direction: an agent using its own legitimately
granted authority on another agent's behalf is, again, an authorized action
occurring at the wrong moment.

\textbf{RC4: defenses are validated against author-constructed attackers
rather than empirically observed real-world failure.} Every isolation and
access-control paper in Section~\ref{sec:taxonomy} reports a defense rate
against attacks the authors themselves constructed. ShellSieve and YoloFS show
what happens when the attacker set instead comes from real deployed artifacts,
1,709 scraped denylists and 290 misuse reports respectively, and find fragility
rates of 69.0\% to 98.6\%~\cite{chen2026shellsieve,zhong2026yolofs}. This is the
same defect Gap 2 (Section~\ref{sec:gaps}) identifies at the level of a single
comparison; RC4 states it as the general pattern behind that gap and behind
Gap~1's absence of shared benchmarks.

\begin{figure*}[!tbp]
\centering
\includegraphics[width=0.92\linewidth]{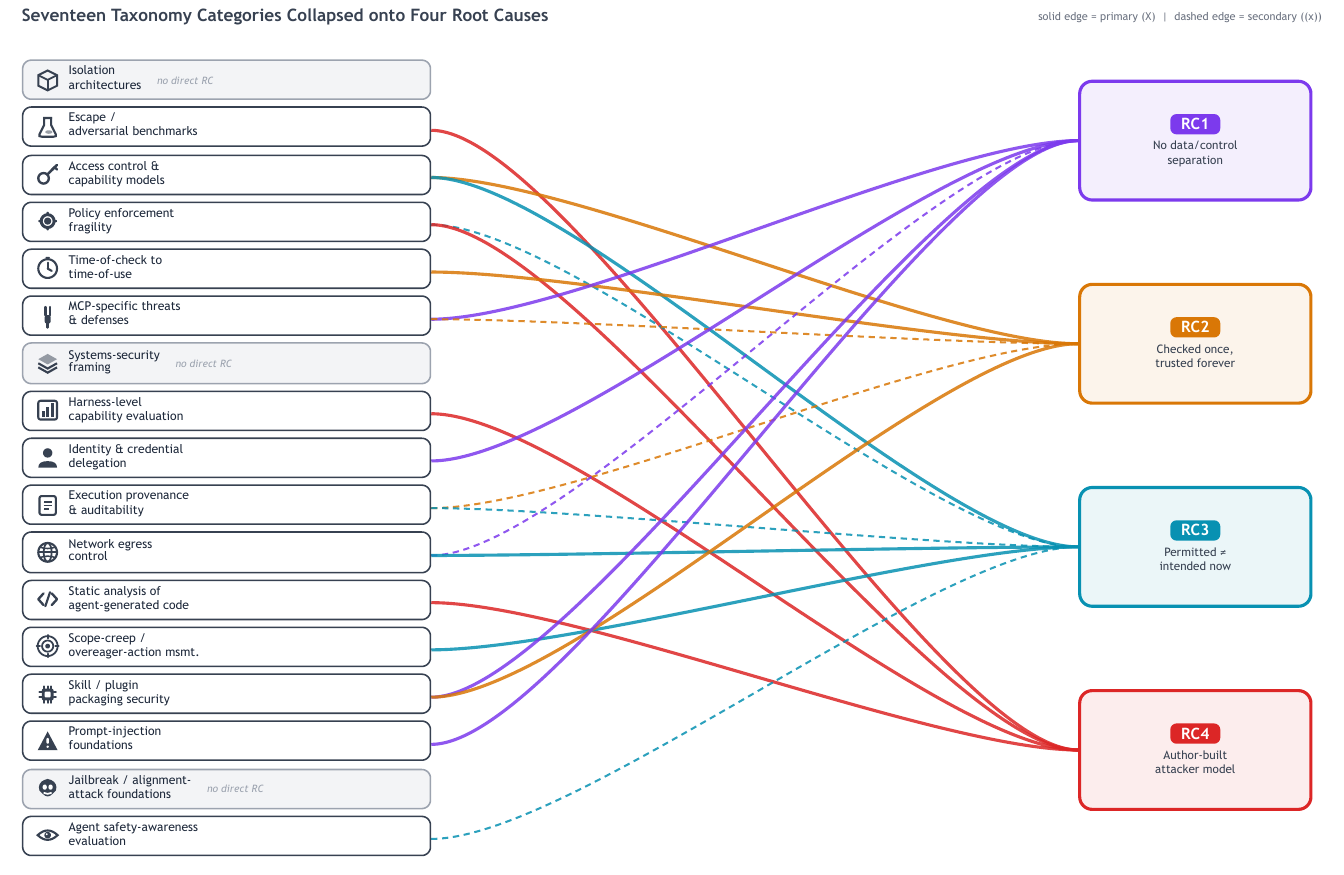}
\caption{Seventeen taxonomy categories collapse onto four recurring root
causes. Dashed edges indicate a category addresses the root cause as a
secondary rather than primary contribution.}
\label{fig:rootcause}
\end{figure*}

Table~\ref{tab:rootcause} maps each of the taxonomy's seventeen categories to
the root cause, or root causes, it is a response to. Six of the seventeen
categories map to more than one root cause, eight map to exactly one, and
three, isolation architectures, systems-security framing, and the
jailbreak/alignment-attack foundations, are not a direct response to any of
the four (a category can still be valuable, as isolation architectures
clearly are, without targeting one of these four specific defects; isolation
bounds the blast radius of an action regardless of which root cause produced
it, which is a different kind of contribution). No category in our corpus is a
response to more than two of the four root causes, meaning no single existing
mechanism claims to, or plausibly could, fix more than half of the recurring
defects we identify.
Figure~\ref{fig:rootcause} presents the same mapping visually: each category is
drawn once, with an edge to every root cause it addresses, so a category with
two edges is one of the nine that targets more than one recurring defect.

\begin{table*}[!tbp]
\centering
\caption{Mapping of taxonomy categories to root causes. A category maps to
RC1--RC4 as defined in Section~\ref{sec:rootcause-four}; a dash means the
category's primary contribution is a mechanism or measurement that does not
target one of the four recurring defects directly (for example, systems-
security framing argues for a change in perspective rather than fixing a
specific defect).}
\label{tab:rootcause}
\footnotesize
\input{tables/tab_rootcause.tex}
\end{table*}

\begin{figure*}[!tbp]
\centering
\includegraphics[width=0.96\linewidth]{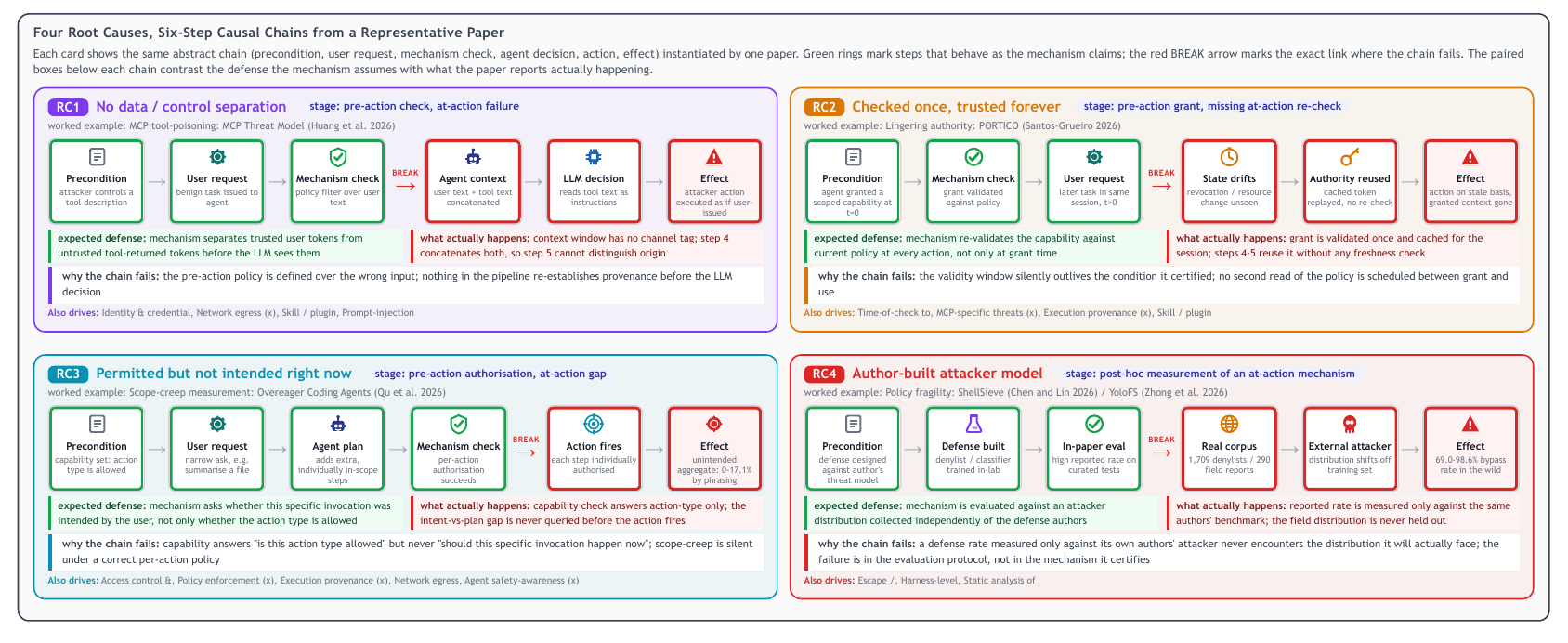}
\caption{The same four causal chains, each instantiated by one representative
paper from the corpus. The break point (red) is a different step in each
chain, showing that ``authorization'' fails for structurally different
reasons across RC1--RC4. Each card also lists the other taxonomy categories
whose techniques map to the same root cause (``Also drives'').}
\label{fig:mechanism}
\end{figure*}

The four root causes are easiest to see side by side as causal chains rather
than as abstract statements, since each is really a claim about which step in
a chain of events breaks down. Figure~\ref{fig:mechanism} instantiates each
root cause with one representative paper from the corpus, tracing the same
four-step shape, precondition, intervening step, action, outcome, and marking
where in that chain the failure actually occurs. Reading the four chains
together makes a pattern in the pattern visible: RC1 and RC2 are failures of
missing state (no tag distinguishing content from instruction; no re-check
step marking when a grant expired), while RC3 and RC4 are failures of an
existing mechanism answering the wrong question (permitted-in-general instead
of intended-right-now; validated against the wrong attacker instead of the
right one). No single existing defense in our corpus addresses both a
missing-state failure and a wrong-question failure at once, which is a second,
independent way of seeing the same ceiling Section~\ref{sec:gaps} identifies
empirically: partial defenses compose in the taxonomy, but they do not compose
along these chains.

\subsection{A pipeline-stage and role reading of the same corpus}
\label{sec:rootcause-pipeline}

A second, orthogonal way to read the corpus is to ask, for each category, where
in an agent's action pipeline its dominant contribution intervenes (before an
action is taken, while it is being taken, or after it has already happened),
and whether the category's dominant contribution is a proposed mechanism or an
empirical measurement of whether an existing mechanism holds. Table~\ref{tab:pipelinestage}
reports this classification. We classify at the category level by each
category's stated primary contribution, consistent with the single-category
assignment described in Section~\ref{sec:method}; several individual papers
within a category combine more than one role, which we already report as a
limitation in Section~\ref{sec:limitations}, so Table~\ref{tab:pipelinestage}
should be read as the dominant pattern rather than an exhaustive per-paper
claim.

\begin{table*}[!tbp]
\centering
\caption{Categories classified by pipeline stage (pre-action, at-action,
post-action) and by role (mechanism vs.\ measurement). Cell entries are the
number of papers in that category.}
\label{tab:pipelinestage}
\footnotesize
\input{tables/tab_pipeline.tex}
\end{table*}

Two patterns stand out. First, five of the six mechanism-proposing categories,
isolation architectures, access control, time-of-check to time-of-use,
identity and credential delegation, and network egress control, intervene
pre-action or at-action, gating a decision before or as it is made, splitting
evenly at eight papers pre-action and eight at-action, while every one of the
eight categories whose dominant contribution is a measurement, escape
benchmarks, enforcement fragility, MCP-specific threats and defenses,
harness-level capability evaluation, static analysis of generated code,
scope-creep measurement, skill and plugin packaging security, and agent
safety-awareness evaluation, evaluates a gate, a threat surface, or a
generated artifact after the fact, by running real or adversarial
trajectories through it, auditing deployed servers or denylists, or scoring
already-produced code, and counting failures. The corpus proposes gates and
separately measures whether gates hold, but, consistent with Gap~1 and Gap~2
(Section~\ref{sec:gaps}), the two groups of papers are not the same papers:
none of the mechanism categories in our corpus report their own defense rate
under the specific adversarial corpora the measurement categories have already
built and released. Second, execution provenance and auditability is the only
post-action mechanism category in the corpus, at one paper, and is
structurally the closest thing to a backstop for RC2 and RC3: a technique
that catches a stale-authorization or scope-mismatch failure after it has
already been enacted cannot prevent the action, but it is the only stage in
the pipeline where such a failure is currently caught at all once a
pre-action gate has already been passed.

\section{Cross-Cutting Gaps}
\label{sec:gaps}

The taxonomy in Section~\ref{sec:taxonomy} is organized the way the field
organizes itself, by mechanism. Reading across categories rather than within
any one surfaces five gaps that are invisible from inside a single category.

\subsection{Gap 1: isolation and access control are not evaluated against each other}
Every isolation architecture we reviewed is evaluated against a fixed attacker
model and reports its own overhead and defense rate. Every access-control or
capability paper does the same, independently, against a different fixed
attacker model. None of the \NIsolationArchitecture{} isolation papers and none
of the \NAccessControl{} access-control papers in our corpus evaluate their
mechanism on a shared benchmark against the other category's mechanism, so a
reader cannot currently learn whether a capability system such as
PORTICO~\cite{santosgrueiro2026portico} or SEAgent~\cite{ji2026seagent} is more
or less effective than an isolation boundary such as
IsolateGPT~\cite{wu2025isolategpt} or ceLLMate~\cite{meng2025cellmate} at
stopping the same attack, or whether the two are complementary layers whose
combination is stronger than either alone. SandboxEscapeBench~\cite{marchand2026sandboxescape}
and RedCode~\cite{guo2024redcode} are the closest things to a shared benchmark
in the corpus, but neither is used, in the papers we surveyed, to compare an
isolation architecture against an access-control system head to head.

\subsection{Gap 2: defenses are not re-evaluated under known enforcement failure rates}
ShellSieve reports that 69.0\% to 98.6\% of real command denylists are
fragile~\cite{chen2026shellsieve}, and YoloFS documents the same
class of failure from 290 real reports~\cite{zhong2026yolofs}. Neither the
isolation architectures nor the access-control papers in our corpus re-evaluate
their own defense under the specific bypass techniques ShellSieve's pipeline
discovers. A defense that reports a strong result against a fixed,
hand-constructed attacker set is not the same claim as a defense that survives
against denylist bypasses empirically shown to defeat the majority of
real-world policies. This gap is at least partially addressable without new
infrastructure: ShellSieve's own bypass corpus is a natural adversarial test set
for any of the access-control papers in Section~\ref{sec:taxonomy}.

\subsection{Gap 3: TOCTOU and MCP threats are studied as separate literatures}
A time-of-check-to-time-of-use race and a tool-poisoning attack on MCP tool
metadata are both, structurally, a validate-then-act sequence in which the
validated state can change before the action executes: TOCTOU work validates a
file or a webpage and then acts on stale content~\cite{lilienthal2025toctou,jiang2026browsertoctou},
while the MCP tool-poisoning threat validates a tool's declared behavior once at
discovery time and then trusts it on every subsequent
call~\cite{huang2026mcpthreat,hasan2025mcpfirstglance}. None of the papers in
either category cites the other, and neither literature uses the other's
vocabulary, even though a defense against one class, re-validating state
immediately before use rather than caching an earlier check, is a candidate
defense against the other. A unified treatment of validate-then-act races across
both settings is, to our knowledge, not yet written.

\subsection{Gap 4: every enforcement mechanism assumes an honest policy author}
The access-control and enforcement papers in our corpus differ in mechanism,
Datalog reference monitors~\cite{palumbo2026forge}, capability
tokens~\cite{santosgrueiro2026portico}, deterministic
pre-action gates~\cite{uchibeke2026oap}, information-flow graphs~\cite{ji2026seagent},
but every one of them assumes the policy itself is correctly specified by a
trustworthy author and asks only whether that policy is then enforced. None
studies what happens when the policy is wrong, overly permissive by mistake,
internally contradictory, or where a policy author under time pressure grants
broader scope than intended because narrower scoping is more work. Given that
ShellSieve's 69--98\% fragility figure is measured against denylists that real
developers wrote and shipped~\cite{chen2026shellsieve}, policy-authoring error
is plausibly at least as large a source of real-world risk as enforcement
failure, and it is the one part of the pipeline this literature does not yet
measure.

\subsection{Gap 5: scope creep is measured but not addressed by any enforcement
mechanism}
The scope-creep measurement in our corpus finds overeager rates on Claude Code
jumping from 0.0\% to 17.1\% depending on prompt phrasing alone, with no change
to the task or the model~\cite{qu2026overeager}. This is not the threat model
any access-control or capability paper in Section~\ref{sec:taxonomy} is built
for: a capability token, a Datalog policy, or a pre-action gate all assume the
dangerous action is either authorized or it is not, but an overeager action is,
by the measurement's own definition, an action the agent was never asked to
take and that a static permission grant would not necessarily forbid, since the
agent may already hold the capability it is misusing for an unrequested
purpose. None of the access-control papers we reviewed cites this measurement,
and the measurement paper does not evaluate whether any of the reviewed
access-control mechanisms would have caught the overeager actions it
documents. A mechanism that constrains \emph{what an agent is allowed to do}
does not, by construction, constrain \emph{whether an authorized capability is
invoked when it should not be}, and this literature does not yet distinguish
the two problems, let alone address the second.

\begin{figure*}[!tbp]
\centering
\includegraphics[width=0.8\linewidth]{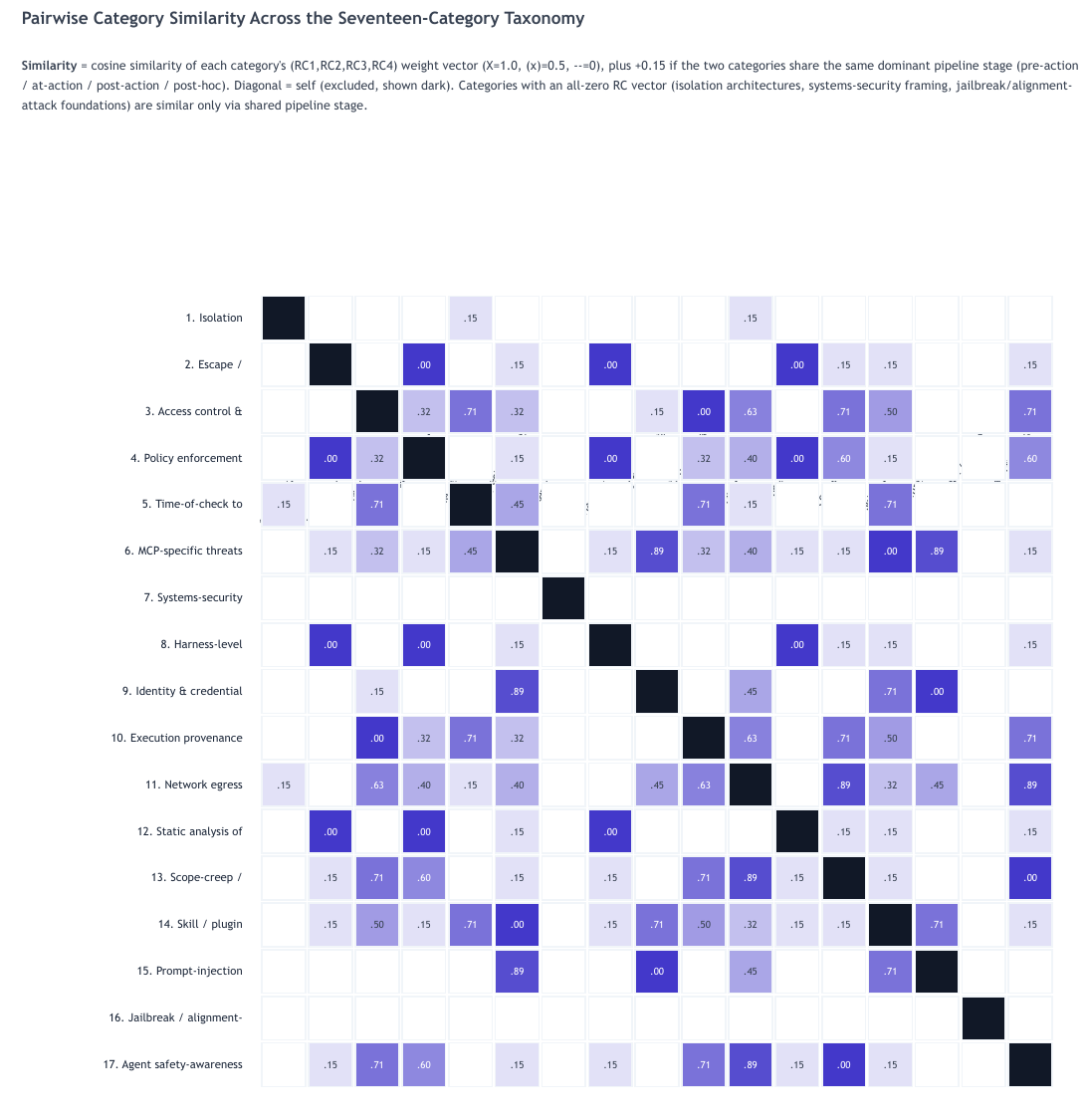}
\caption{Pairwise similarity across the seventeen taxonomy categories, scored
by cosine similarity of each category's root-cause weight vector plus a bonus
for sharing a dominant pipeline stage. High off-diagonal scores concentrate on
category pairs that Gaps~1--3 argue are studied in isolation despite
addressing structurally related failures.}
\label{fig:heatmap}
\end{figure*}

The five gaps above are easiest to state one at a time, but they share a
common structural cause: the taxonomy in Section~\ref{sec:taxonomy} groups
papers by the mechanism a paper proposes, and that grouping systematically
separates categories that are, by the root-cause and pipeline-stage readings
of Section~\ref{sec:rootcause}, close cousins. Figure~\ref{fig:heatmap} makes
this concrete by scoring every pair of the seventeen categories on shared root
causes and shared pipeline stage. Several of the highest off-diagonal scores
connect exactly the categories Gaps~1--3 argue should already be in dialogue,
isolation architectures and access control (Gap~1), enforcement fragility and
MCP-specific threats (Gap~2), and time-of-check-to-time-of-use and MCP-specific
threats (Gap~3), yet none of the underlying papers cites across that pair. The
heatmap does not show a corpus that is randomly fragmented; it shows a corpus
whose highest-similarity pairs are consistently the ones the field has not yet
connected, which is a second, independent way of observing the same gaps
Sections~\ref{sec:gaps}.1--\ref{sec:gaps}.3 identify by close reading.

\section{Related Work}
\label{sec:related}

The closest related work is four broader agentic-AI-security surveys we
excluded from the taxonomy itself because each is a secondary synthesis of the
field rather than a primary execution-security contribution.
\cite{dehghantanha2026attacksurface} organizes over 20 studies into a
four-branch attack taxonomy (prompt-level injection, knowledge-base poisoning,
tool and plug-in exploits, multi-agent emergent threats) and reviews sandboxes
as one item within a broader defense category alongside input sanitization and
access control. \cite{shahriar2025agenticsecurity} reviews over 260 papers
across three pillars, applications, threats, and defenses, spanning the full
agent lifecycle and the use of agents for offensive and defensive security
work, with execution isolation appearing as one defense among many rather than
as a dedicated subject. \cite{maloyan2026promptinjectionsok} is itself
titled as an SoK and synthesizes 78 studies into a three-part taxonomy of
delivery vectors, attack modalities, and propagation behaviors specifically for
prompt injection against coding assistants, cataloging 42 attack techniques and
18 defenses; sandboxing appears as one architectural concern feeding its
proposed defense-in-depth framework rather than as its own category, and its
scope, prompt injection specifically, is narrower than and largely orthogonal
to execution security, since a perfectly sandboxed agent can still be
successfully injected, and a perfectly resistant model still needs a sandbox if
it is ever compromised. \cite{chen2026claweddangerous} reviews 50 papers on
open agentic platforms and proposes a six-dimensional analytical taxonomy and a
platform security scoring system, concluding the field has strong coverage of
attack cataloging and benchmarks but weak coverage of deployment controls,
operational governance, and capability revocation, a conclusion this paper's
taxonomy independently supports from a different literature sample. All four
surveys are valuable for the breadth they cover, applications, multi-agent
risk, prompt-level attacks, platform governance, that this paper does not
attempt; the gap they leave, and the one this paper is written to fill, is a
systematization of execution security specifically, with enough granularity to
compare isolation architectures against capability models, benchmarks against
the defenses they are meant to stress-test, and enforcement mechanisms against
the empirical fragility rates measured against real deployed policies.

A second and fast-moving line of recent work develops defenses at the
\emph{injection} layer, aiming to stop a malicious instruction from ever
reaching the model in the first place. MELON detects indirect prompt injection
by re-executing the agent loop with masked tool results and flagging behavioral
divergence \cite{zhu2025melon}; PromptArmor screens tool outputs for injected
instructions before they re-enter the context window
\cite{shi2025promptarmor}; and ARGUS conditions its detector on the task
context to catch instructions that are only adversarial relative to the current
goal \cite{weng2026argus}. These 2025--2026 mechanisms are complementary to,
but deliberately outside, our scope: they target the boundary at which a model
is compromised, whereas our corpus systematizes the execution-layer containment
that must hold \emph{after} compromise, on the premise, reinforced by a recent
result that no injection defense achieves perfect recall under adaptive
pressure \cite{abdelnabi2026alwaysfall}, that a residual injection rate must be
assumed rather than engineered to zero. The two layers are independent axes of
the same problem, and a deployed coding agent needs both; we treat injection
defenses as the upstream complement to, not a substitute for, the sandboxing,
capability, provenance, and policy-enforcement mechanisms catalogued here.

\section{Discussion}
\label{sec:discussion}

The volume of work in this space, \NCorpus{} papers meeting our inclusion
criteria in under two years, indicates the field recognizes execution security
as a real and urgent problem. Table~\ref{tab:years} breaks this corpus down by
publication year and shows the growth is not linear: the majority of the
corpus was published in the most recent year alone, consistent with execution
security emerging as an active concern only after coding agents began
operating with broad, largely unsandboxed tool access.

\begin{table}[h]
\centering
\caption{Corpus size by publication year.}
\label{tab:years}
\input{tables/tab_years.tex}
\end{table}

The four disclosed CVEs in
Section~\ref{sec:incidents} and the OWASP Top 10's inclusion of excessive
agency as a named risk predate or overlap most of this corpus and show the
underlying concern is not speculative~\cite{cve202421626,cve202553773,cve202559536,cve202621852,owasp2025llmtop10}.
What the taxonomy in Section~\ref{sec:taxonomy} shows is that the field's
energy has gone almost entirely into building new mechanisms, five isolation
architectures, six access-control frameworks, four MCP-specific defenses, two
identity-delegation proposals, and comparatively little into cross-mechanism
evaluation. This is a familiar pattern in a young subfield: mechanism proposals
are easier to publish than the harder comparative and adversarial evaluation
work that would tell a practitioner which mechanism to actually deploy, or
whether to deploy more than one. The newer categories in our corpus, execution
provenance, network egress control, static analysis of generated code,
scope-creep measurement, and skill packaging security, each have exactly one
paper, which is either a sign these are genuinely nascent directions or a sign
our search has not yet found the rest of a larger literature; we treat this as
an open question the released corpus file is meant to help answer as more work
appears.

The two enforcement-fragility papers in our corpus are, in our reading, the
most consequential finding in the whole literature, precisely because they
measure real, shipped artifacts rather than a mechanism's own claimed
guarantee. A defense that has not been tested against ShellSieve's bypass
corpus~\cite{chen2026shellsieve} or against the failure patterns YoloFS
documents~\cite{zhong2026yolofs} has not yet been tested against the kind of
policy a real team would actually write under time pressure. We would
encourage authors of future isolation and access-control papers to treat these
two papers' adversarial corpora as a standard evaluation baseline, in the same
way image classification adopted common corruption benchmarks once they
existed, rather than evaluating only against attacks the authors constructed
themselves.

\section{Limitations}
\label{sec:limitations}

This is a systematization, not a measurement study, and its central limitation
is coverage rather than measurement error. Our search was iterative and
citation-following rather than exhaustive, and arXiv posting norms in this area
mean some relevant work may not surface under the search terms we used, or may
have been posted after this paper was written; the released corpus file is
intended to make later extension straightforward rather than to claim
completeness. The categorization itself involves judgment: several papers,
notably SEAgent~\cite{ji2026seagent} and FORGE~\cite{palumbo2026forge}, combine
elements of more than one category, and we assign each to the single category
that best matches its primary contribution rather than allowing multiple
category membership, which understates the degree of overlap between
categories. Eight of our seventeen categories currently contain exactly one paper
each (identity delegation and four others contain two); we report this honestly in
Section~\ref{sec:discussion} as ambiguous between a genuinely nascent research
direction and an incomplete search, and a reader should weight claims about
those categories accordingly. The verification protocol confirms that a
paper's title, authors, and venue match its abstract page and catches gross
misattribution, of the kind we found and corrected for two candidates during
this review, but it does not constitute independent replication of any paper's
empirical claims; we report those claims as stated by their authors. The four
CVEs in Section~\ref{sec:incidents} were confirmed directly against NVD listing
pages, which we treat as authoritative for the affected product, version, and
CVSS score, but we did not independently reproduce any of the underlying
exploits. Finally, the five gaps in Section~\ref{sec:gaps} are argued from the
absence of a citation or a shared benchmark across categories, which is a real
but indirect form of evidence; it is possible relevant cross-category work
exists in a venue or preprint outside the corpus we searched.

\section{Future Work}
\label{sec:future}

The most direct next step is closing Gap 1 empirically: running a small number
of representative isolation architectures and access-control mechanisms against
a shared attacker corpus, such as SandboxEscapeBench's introduced
weaknesses~\cite{marchand2026sandboxescape} or RedCode's risky-operation
suite~\cite{guo2024redcode}, and reporting whether the two categories of
defense are substitutes or complements. Closing Gap 2 requires no new
infrastructure at all: re-running ShellSieve's released bypass
corpus~\cite{chen2026shellsieve} against any of the access-control mechanisms in
Section~\ref{sec:taxonomy} is a direct, low-cost evaluation that would turn a
mechanism's untested claim into a measured one. Gap 3 calls for a unified formal
treatment of validate-then-act races that covers both the filesystem and
browser-state TOCTOU settings and the MCP tool-metadata setting under one model,
so that a mitigation designed for one setting can be checked against the other
without translation. Gap 4 is, in our view, the most consequential of the first four: an empirical
study of policy-authoring error itself, in the style of ShellSieve's denylist
study but aimed at the access-control policies proposed in
Section~\ref{sec:taxonomy} rather than at command denylists, would tell the
field whether the mechanisms it has built are being undermined more by weak
enforcement or by policies that were never correctly specified in the first
place. Gap 5 calls for the reverse experiment to the one OverEager-Bench
already ran~\cite{qu2026overeager}: taking the overeager trajectories that
benchmark has already collected and replaying them against a capability system
such as PORTICO~\cite{santosgrueiro2026portico} or an information-flow
framework such as SEAgent~\cite{ji2026seagent} to measure directly what
fraction of scope creep an authorization mechanism built for a different
threat model happens to catch, rather than assuming from the mechanism's
design alone that it would or would not.

\section{Conclusion}
\label{sec:conclusion}

We systematized \NCorpus{} papers on AI coding agent execution security,
published between \YearMin{} and \YearMax{}, into \NCategories{} verified
categories, correcting two misattributions found during verification along the
way and independently confirming four disclosed, patched CVEs, two of them in
Claude Code itself, directly against the NIST National Vulnerability Database.
Four existing broader surveys of agentic AI security discuss sandboxing only as
one entry among many defenses, or treat prompt injection specifically rather
than execution security generally; this paper is the first, to our knowledge,
to treat execution security as a subject with its own internal structure.
Reading across the taxonomy rather than within any single category surfaces
five gaps that individual papers cannot see from inside their own category:
isolation and access control are evaluated in isolation from each other,
defenses are not re-tested against measured real-world enforcement failure
rates, TOCTOU and MCP threats are treated as unrelated despite sharing a common
structure, every enforcement mechanism in the literature assumes a policy
author who does not make mistakes, and a newly measured failure mode, benign
scope creep at rates up to 17.1\%, is addressed by no access-control mechanism
in the corpus. Closing these gaps, not proposing another isolation mechanism or
access-control model, is where we believe the field's next real progress lies.

\bibliographystyle{IEEEtran}
\bibliography{bibliography}

\appendices

\section{Per-Paper Classification of the Corpus}
\label{app:corpus}

Table~\ref{tab:appendix-corpus} lists every one of the \NCorpus{} papers that
enter the taxonomy, together with its category, root-cause attribution, and
pipeline-stage role, so that the taxonomy in Section~\ref{sec:taxonomy} can be
audited paper by paper rather than only at the category level.  The row order
inside each category is by publication year, then by citation key.  Role
codes are: \emph{mech.} for a mechanism proposal, \emph{meas.} for a
measurement or empirical evaluation, \emph{found.} for a foundational
threat-vocabulary paper without a specific pipeline-stage contribution, and
\emph{persp.} for a systems-security framing paper.  Together with
Table~\ref{tab:rootcause} (Section~\ref{sec:rootcause}) and
Table~\ref{tab:pipelinestage} (Section~\ref{sec:rootcause-pipeline}), this
table is what makes the reader able to reproduce the corpus and challenge
individual assignments without having to re-read the whole survey.

\footnotesize
\input{tables/tab_appendix_corpus.tex}
\normalsize

\section{Verification Protocol}
\label{app:verification}

The verification protocol referenced from Section~\ref{sec:method} was applied
uniformly to every paper in the corpus.  For each candidate paper we
independently fetched its own abstract page (arXiv preprint page or venue
proceedings entry) and recorded the paper's own claimed title, authors,
publication year, venue, and one-sentence contribution.  A candidate was
admitted to the corpus only when these five fields agreed with the
citation entry we had drafted from a secondary source; disagreements were
resolved in favour of the primary source and the citation entry was rewritten
before admission.  The four CVE incidents reported in
Section~\ref{sec:incidents} were verified separately, directly against the
NIST National Vulnerability Database, and are counted only after the
CVE-ID, affected product, affected version range, and disclosure date agreed
with the vendor's own advisory.  The full verification pipeline (a script
that walks the corpus file, fetches each paper's abstract page, parses its
metadata, and reports mismatches) is released together with the paper as part
of the supplementary artifact, so any reader can re-run the check on their
own machine.  During
verification two misattributions were caught and corrected: one paper had
been listed under isolation architectures on the basis of a secondary
summary but its abstract page describes a policy-enforcement measurement
study, and one entry conflated two separate MCP papers by the same first
author.  Both corrections are reflected in the counts reported in the body
of the paper.

\section{Extended Worked Examples per Root Cause}
\label{app:worked-examples}

Figure~\ref{fig:mechanism} shows one worked example per root cause.  For
completeness, we list here the additional papers in the corpus whose causal
chains match each root-cause pattern, together with the specific step at
which the chain breaks.  These are not new attacks; they are the same
break-points documented in the papers' own evaluation sections, re-stated in
the six-step (precondition, user request, mechanism check, agent decision,
action, effect) framing used in Figure~\ref{fig:mechanism} so that they
compose with the rest of the taxonomy.

\subsection*{Additional examples for RC1 (no data / control separation)}
The tool-poisoning pattern from Figure~\ref{fig:mechanism} recurs, with the
same break between the mechanism-check and agent-context steps, in
skill-package foundations~\cite{li2026secureskills}, the identity-delegation
setting where a delegated token carries data the delegator did not
intend~\cite{south2025delegation}, and the prompt-injection foundations
literature that first isolated the mixing problem in a non-agentic
setting~\cite{greshake2023indirect,liu2023promptinjapps}.  Network egress
control~\cite{abdelnabi2025firewalls} is a secondary contributor because it
rebuilds the separation from the outbound side rather than at ingest.

\subsection*{Additional examples for RC2 (checked once, trusted forever)}
The same stale-check pattern also appears in the TOCTOU
literature~\cite{lilienthal2025toctou,jiang2026browsertoctou}, where the
mechanism check is even earlier in the chain (at file-open or DOM-read time)
and the state that drifts is the filesystem or DOM rather than the policy
snapshot.  Provenance and auditability
work~\cite{sequeira2026agentsentry} contributes secondarily by making the
missed re-check observable after the fact, which is how many of these bugs
are first noticed rather than prevented.  Skill and plugin
packaging~\cite{li2026secureskills} straddles RC1 and RC2 because the
package signature is checked at install time and reused across every later
invocation without a re-check.

\subsection*{Additional examples for RC3 (permitted but not intended now)}
Beyond scope-creep measurement~\cite{qu2026overeager}, RC3 is the dominant
pattern in access-control and capability
work~\cite{buhler2025agentbound,santosgrueiro2026portico,ji2026seagent,li2025aac,uchibeke2026oap,palumbo2026forge},
each of which grants an action type and then relies on the agent's own plan
to stay within intent.  Network egress control~\cite{abdelnabi2025firewalls}
also lives here when the destination is permitted but the specific
per-request payload was not intended by the user.  Agent safety-awareness
evaluations~\cite{yuan2024rjudge} contribute secondarily by asking whether
the model itself can flag such invocations before they fire, which if
achievable would let a per-action mechanism address the gap that a
per-type mechanism cannot.

\subsection*{Additional examples for RC4 (author-built attacker model)}
The pattern from Figure~\ref{fig:mechanism} for policy-enforcement
fragility~\cite{chen2026shellsieve,zhong2026yolofs} also drives escape and
adversarial benchmarks~\cite{marchand2026sandboxescape,guo2024redcode,rabin2025sandboxeval,ruan2023toolemu,zhan2024injecagent,debenedetti2024agentdojo},
harness-level capability evaluation~\cite{mayoralvilches2026csi}, and static
analysis of agent-generated code~\cite{chen2025securevibebench}.  In each
case the reported defense or detection rate is measured against a corpus
generated by the same authors under a threat model of their own choosing;
the failure is a property of the evaluation protocol rather than of the
mechanism itself, which is why it composes with mechanisms that are
otherwise unrelated (isolation, capability, static analysis).

\section{Reproducibility Notes}
\label{app:reproducibility}

The corpus, taxonomy, and all body figures and tables are generated from a
single machine-readable data file by two generator scripts, one for the
tables and one for the figures, plus a verifier that re-derives every
category count, every root-cause count, and every pipeline-stage count from
the data file and checks them against the numbers used in the body of the
paper.  All four artifacts, the data file, the two generators, and the
verifier, are released with the paper as a supplementary artifact.  The
verifier takes no arguments and reports a single pass/fail summary, so any
change to the corpus that would silently drift the body numbers is caught
before the paper builds.  The paper itself builds with a standard
\LaTeX{}/BibTeX cycle and produces a warning-free log on the current corpus.

\end{document}

%% file: tables/macros.tex
\newcommand{\NCorpus}{39}
\newcommand{\NCategories}{17}
\newcommand{\YearMin}{2023}
\newcommand{\YearMax}{2026}
\newcommand{\NIsolationArchitecture}{5}
\newcommand{\NEscapeBenchmark}{6}
\newcommand{\NAccessControl}{6}
\newcommand{\NEnforcementFragility}{2}
\newcommand{\NToctou}{2}
\newcommand{\NMcpSpecific}{4}
\newcommand{\NSystemsFraming}{2}
\newcommand{\NHarnessCapability}{1}
\newcommand{\NIdentityDelegation}{2}
\newcommand{\NExecutionProvenance}{1}
\newcommand{\NNetworkEgressControl}{1}
\newcommand{\NStaticAnalysisGeneratedCode}{1}
\newcommand{\NScopeCreepMeasurement}{1}
\newcommand{\NSkillArchitecture}{1}

%% file: tables/tab_taxonomy.tex
\begin{tabularx}{\textwidth}{@{}lcX@{}}
\toprule
Category & Papers & Representative work \\
\midrule
Isolation architectures & 5 & \cite{wu2025isolategpt}, \cite{meng2025cellmate}, \cite{piao2025agentbay}, \cite{dong2026deltabox}, \cite{yan2025faulttolerant} \\
Escape and adversarial benchmarks & 6 & \cite{marchand2026sandboxescape}, \cite{guo2024redcode}, \cite{rabin2025sandboxeval}, \cite{ruan2023toolemu}, \cite{zhan2024injecagent}, \cite{debenedetti2024agentdojo} \\
Access control and capability models & 6 & \cite{buhler2025agentbound}, \cite{santosgrueiro2026portico}, \cite{ji2026seagent}, \cite{li2025aac}, \cite{uchibeke2026oap}, \cite{palumbo2026forge} \\
Policy enforcement fragility & 2 & \cite{chen2026shellsieve}, \cite{zhong2026yolofs} \\
Time-of-check to time-of-use & 2 & \cite{lilienthal2025toctou}, \cite{jiang2026browsertoctou} \\
MCP-specific threats and defenses & 4 & \cite{huang2026mcpthreat}, \cite{hou2025mcplandscape}, \cite{hasan2025mcpfirstglance}, \cite{xing2026mcpguard} \\
Systems-security framing & 2 & \cite{christodorescu2025systemssecurity}, \cite{wang2026hciframing} \\
Harness-level capability evaluation & 1 & \cite{mayoralvilches2026csi} \\
Identity and credential delegation & 2 & \cite{south2025delegation}, \cite{zhou2026capabilitybinding} \\
Execution provenance and auditability & 1 & \cite{sequeira2026agentsentry} \\
Network egress control & 1 & \cite{abdelnabi2025firewalls} \\
Static analysis of agent-generated code & 1 & \cite{chen2025securevibebench} \\
Scope-creep and overeager-action measurement & 1 & \cite{qu2026overeager} \\
Skill and plugin packaging security & 1 & \cite{li2026secureskills} \\
Prompt-injection foundations & 2 & \cite{greshake2023indirect}, \cite{liu2023promptinjapps} \\
Jailbreak and alignment-attack foundations & 1 & \cite{zou2023universal} \\
Agent safety-awareness evaluation & 1 & \cite{yuan2024rjudge} \\
\midrule
\textbf{Total} & \textbf{39} & \\
\bottomrule
\end{tabularx}

%% file: tables/tab_rootcause.tex
\begin{tabularx}{\textwidth}{@{}X*{4}{>{\centering\arraybackslash}p{1.55cm}}@{}}
\toprule
Category & RC1: no data/ control separation & RC2: checked once, trusted forever & RC3: permitted but not intended now & RC4: author-built attacker model \\
\midrule
Isolation architectures                          & --  & --  & --  & --  \\
Escape and adversarial benchmarks                & --  & --  & --  & X   \\
Access control and capability models             & --  & X   & X   & --  \\
Policy enforcement fragility                     & --  & --  & (x) & X   \\
Time-of-check to time-of-use                     & --  & X   & --  & --  \\
MCP-specific threats and defenses                & X   & (x) & --  & --  \\
Systems-security framing                         & --  & --  & --  & --  \\
Harness-level capability evaluation              & --  & --  & --  & X   \\
Identity and credential delegation               & X   & --  & --  & --  \\
Execution provenance and auditability            & --  & (x) & (x) & --  \\
Network egress control                           & (x) & --  & X   & --  \\
Static analysis of agent-generated code          & --  & --  & --  & X   \\
Scope-creep and overeager-action measurement     & --  & --  & X   & --  \\
Skill and plugin packaging security              & X   & X   & --  & --  \\
Prompt-injection foundations                     & X   & --  & --  & --  \\
Jailbreak and alignment-attack foundations       & --  & --  & --  & --  \\
Agent safety-awareness evaluation                & --  & --  & (x) & --  \\
\bottomrule
\end{tabularx}

%% file: tables/tab_pipeline.tex
\begin{tabularx}{\textwidth}{@{}X*{3}{>{\centering\arraybackslash}p{2.2cm}}@{}}
\toprule
Category & Papers & Pipeline stage & Role \\
\midrule
Isolation architectures                          & 5 & At-action    & Mechanism   \\
Escape and adversarial benchmarks                & 6 & Post-hoc     & Measurement \\
Access control and capability models             & 6 & Pre-action   & Mechanism   \\
Policy enforcement fragility                     & 2 & Post-hoc     & Measurement \\
Time-of-check to time-of-use                     & 2 & At-action    & Mechanism   \\
MCP-specific threats and defenses                & 4 & Post-hoc     & Measurement \\
Systems-security framing                         & 2 & --           & Perspective \\
Harness-level capability evaluation              & 1 & Post-hoc     & Measurement \\
Identity and credential delegation               & 2 & Pre-action   & Mechanism   \\
Execution provenance and auditability            & 1 & Post-action  & Mechanism   \\
Network egress control                           & 1 & At-action    & Mechanism   \\
Static analysis of agent-generated code          & 1 & Post-hoc     & Measurement \\
Scope-creep and overeager-action measurement     & 1 & Post-hoc     & Measurement \\
Skill and plugin packaging security              & 1 & Post-hoc     & Measurement \\
Prompt-injection foundations                     & 2 & --           & Foundational \\
Jailbreak and alignment-attack foundations       & 1 & --           & Foundational \\
Agent safety-awareness evaluation                & 1 & Post-hoc     & Measurement \\
\midrule
\textbf{Total} & \textbf{39} & & \\
\bottomrule
\end{tabularx}

%% file: tables/tab_years.tex
\begin{tabular}{@{}lc@{}}
\toprule
Year & Papers \\
\midrule
2023 & 4 \\
2024 & 4 \\
2025 & 14 \\
2026 & 17 \\
\bottomrule
\end{tabular}

%% file: tables/tab_appendix_corpus.tex
\begin{table*}[!tbp]
\centering
\caption{Per-paper classification of the \NCorpus{}-paper corpus.  RC assignments follow Table~\ref{tab:rootcause}; pipeline stage follows Table~\ref{tab:pipelinestage}.  Parenthesised RC entries denote secondary attribution.  Role: \emph{mech.}\ mechanism proposal, \emph{meas.}\ measurement/evaluation, \emph{found.}\ foundational threat vocabulary, \emph{persp.}\ systems-security framing.}
\label{tab:appendix-corpus}
\footnotesize
\begin{tabular}{@{}rllllllr@{}}
\toprule
\# & First author, year & Venue & arXiv~ID & Category & RC & Stage & Role \\
\midrule
1 & Luoxi Meng, 2025 & arXiv & 2512.12594 & Isolation architectures & -- & at-action & mech. \\
2 & Yun Piao, 2025 & arXiv & 2512.04367 & Isolation architectures & -- & at-action & mech. \\
3 & Yuhao Wu, 2025 & NDSS & 2403.04960 & Isolation architectures & -- & at-action & mech. \\
4 & Boyang Yan, 2025 & arXiv & 2512.12806 & Isolation architectures & -- & at-action & mech. \\
5 & Yunpeng Dong, 2026 & arXiv & 2605.22781 & Isolation architectures & -- & at-action & mech. \\
6 & Yangjun Ruan, 2023 & ICLR & 2309.15817 & Escape/adversarial benchmarks & RC4 & post-hoc & meas. \\
7 & Edoardo Debenedetti, 2024 & NeurIPS D\&B & 2406.13352 & Escape/adversarial benchmarks & RC4 & post-hoc & meas. \\
8 & Chengquan Guo, 2024 & NeurIPS & 2411.07781 & Escape/adversarial benchmarks & RC4 & post-hoc & meas. \\
9 & Qiusi Zhan, 2024 & ACL Findings & 2403.02691 & Escape/adversarial benchmarks & RC4 & post-hoc & meas. \\
10 & Rafiqul Rabin, 2025 & arXiv & 2504.00018 & Escape/adversarial benchmarks & RC4 & post-hoc & meas. \\
11 & Rahul Marchand, 2026 & arXiv & 2603.02277 & Escape/adversarial benchmarks & RC4 & post-hoc & meas. \\
12 & Christoph Buhler, 2025 & arXiv & 2510.21236 & Access control \& capability & RC2, RC3 & pre-action & mech. \\
13 & Xinfeng Li, 2025 & arXiv & 2510.11108 & Access control \& capability & RC2, RC3 & pre-action & mech. \\
14 & Zimo Ji, 2026 & arXiv & 2601.11893 & Access control \& capability & RC2, RC3 & pre-action & mech. \\
15 & Nils Palumbo, 2026 & arXiv & 2602.16708 & Access control \& capability & RC2, RC3 & pre-action & mech. \\
16 & Igor Santos-Grueiro, 2026 & arXiv & 2606.22504 & Access control \& capability & RC2, RC3 & pre-action & mech. \\
17 & Uchi Uchibeke, 2026 & arXiv & 2603.20953 & Access control \& capability & RC2, RC3 & pre-action & mech. \\
18 & Chuyang Chen, 2026 & arXiv & 2606.15549 & Policy enforcement fragility & RC4 (RC3) & post-hoc & meas. \\
19 & Shawn Wanxiang Zhong, 2026 & arXiv & 2604.13536 & Policy enforcement fragility & RC4 (RC3) & post-hoc & meas. \\
20 & Derek Lilienthal, 2025 & arXiv & 2508.17155 & TOCTOU & RC2 & at-action & mech. \\
21 & Linxi Jiang, 2026 & arXiv & 2603.00476 & TOCTOU & RC2 & at-action & mech. \\
22 & Mohammed Mehedi Hasan, 2025 & arXiv & 2506.13538 & MCP-specific & RC1 (RC2) & post-hoc & meas. \\
23 & Xinyi Hou, 2025 & arXiv & 2503.23278 & MCP-specific & RC1 (RC2) & post-hoc & meas. \\
24 & Charoes Huang, 2026 & arXiv & 2603.22489 & MCP-specific & RC1 (RC2) & post-hoc & meas. \\
25 & Wenpeng Xing, 2026 & arXiv & 2508.10991 & MCP-specific & RC1 (RC2) & post-hoc & meas. \\
26 & Mihai Christodorescu, 2025 & arXiv & 2512.01295 & Systems-security framing & -- & -- & persp. \\
27 & Peiran Wang, 2026 & arXiv & 2605.24309 & Systems-security framing & -- & -- & persp. \\
28 & Victor Mayoral-Vilches, 2026 & arXiv & 2605.28334 & Harness-level capability eval. & RC4 & post-hoc & meas. \\
29 & Tobin South, 2025 & arXiv & 2501.09674 & Identity/credential delegation & RC1 & pre-action & mech. \\
30 & Ziling Zhou, 2026 & arXiv & 2603.14332 & Identity/credential delegation & RC1 & pre-action & mech. \\
31 & Rohan Sequeira, 2026 & arXiv & 2603.22868 & Execution provenance & (RC2, RC3) & post-action & mech. \\
32 & Sahar Abdelnabi, 2025 & arXiv & 2502.01822 & Network egress control & RC3 (RC1) & at-action & mech. \\
33 & Junkai Chen, 2025 & arXiv & 2509.22097 & Static analysis of gen. code & RC4 & post-hoc & meas. \\
34 & Yubin Qu, 2026 & arXiv & 2605.18583 & Scope-creep measurement & RC3 & post-hoc & meas. \\
35 & Zhiyuan Li, 2026 & arXiv & 2604.02837 & Skill/plugin packaging & RC1, RC2 & post-hoc & meas. \\
36 & Kai Greshake, 2023 & AISec & 2302.12173 & Prompt-injection foundations & RC1 & -- & found. \\
37 & Yi Liu, 2023 & arXiv & 2306.05499 & Prompt-injection foundations & RC1 & -- & found. \\
38 & Andy Zou, 2023 & arXiv & 2307.15043 & Jailbreak/alignment foundations & -- & -- & found. \\
39 & Tongxin Yuan, 2024 & EMNLP Findings & 2401.10019 & Agent safety-awareness eval. & (RC3) & post-hoc & meas. \\
\bottomrule
\end{tabular}
\end{table*}

%% file: Paper.bbl
\begin{thebibliography}{10}
\providecommand{\url}[1]{#1}
\csname url@samestyle\endcsname
\providecommand{\newblock}{\relax}
\providecommand{\bibinfo}[2]{#2}
\providecommand{\BIBentrySTDinterwordspacing}{\spaceskip=0pt\relax}
\providecommand{\BIBentryALTinterwordstretchfactor}{4}
\providecommand{\BIBentryALTinterwordspacing}{\spaceskip=\fontdimen2\font plus
\BIBentryALTinterwordstretchfactor\fontdimen3\font minus
  \fontdimen4\font\relax}
\providecommand{\BIBforeignlanguage}[2]{{%
\expandafter\ifx\csname l@#1\endcsname\relax
\typeout{** WARNING: IEEEtran.bst: No hyphenation pattern has been}%
\typeout{** loaded for the language `#1'. Using the pattern for}%
\typeout{** the default language instead.}%
\else
\language=\csname l@#1\endcsname
\fi
#2}}
\providecommand{\BIBdecl}{\relax}
\BIBdecl

\bibitem{dehghantanha2026attacksurface}
A.~Dehghantanha and S.~Homayoun, ``{SoK}: The attack surface of agentic {AI},
  tools, and autonomy,'' \emph{arXiv preprint arXiv:2603.22928}, 2026.

\bibitem{shahriar2025agenticsecurity}
A.~Shahriar, M.~N. Rahman, S.~Ahmed, F.~Sadeque, and M.~R. Parvez, ``A survey
  on agentic security: Applications, threats and defenses,'' \emph{arXiv
  preprint arXiv:2510.06445}, 2025.

\bibitem{maloyan2026promptinjectionsok}
N.~Maloyan and D.~Namiot, ``Prompt injection attacks on agentic coding
  assistants: A systematic analysis of vulnerabilities in skills, tools, and
  protocol ecosystems,'' \emph{arXiv preprint arXiv:2601.17548}, 2026.

\bibitem{chen2026claweddangerous}
S.~Chen, Q.~Wang, G.~Yu, X.~Wang, and L.~Zhu, ``Clawed and dangerous: Can we
  trust open agentic systems?'' \emph{arXiv preprint arXiv:2603.26221}, 2026.

\bibitem{cve202421626}
{NIST National Vulnerability Database}, ``{CVE-2024-21626}: runc container
  escape via leaked file descriptor referencing the host working directory,''
  MITRE / NIST NVD, Tech. Rep., 2024, ``Leaky Vessels''; also assigned
  GHSA-xr7r-f8xq-vfvv.

\bibitem{cve202553773}
------, ``{CVE-2025-53773}: Command injection in {GitHub} copilot and visual
  studio,'' MITRE / NIST NVD, Tech. Rep., 2025, cVSS 3.1 base score 7.8 (High);
  affects Visual Studio 2022 17.14.0 to 17.14.11.

\bibitem{cve202559536}
------, ``{CVE-2025-59536}: Code injection in {Claude Code} via startup trust
  dialog bypass,'' MITRE / NIST NVD, Tech. Rep., 2025, cVSS 3.1 base score 8.8
  (High); CWE-94; fixed in Claude Code 1.0.111; GHSA-4fgq-fpq9-mr3g.

\bibitem{cve202621852}
------, ``{CVE-2026-21852}: Data exfiltration in {Claude Code} project-load
  flow before trust confirmation,'' MITRE / NIST NVD, Tech. Rep., 2026, cVSS
  3.1 base score 7.5 (High); fixed in Claude Code 2.0.65.

\bibitem{christodorescu2025systemssecurity}
M.~Christodorescu, E.~Fernandes, A.~Hooda, S.~Jha, J.~Rehberger, K.~Chaudhuri,
  X.~Fu, K.~Shams, G.~Amir, J.~Choi, S.~Choudhary, N.~Palumbo, A.~Labunets, and
  N.~V. Pandya, ``Systems security foundations for agentic computing,''
  \emph{arXiv preprint arXiv:2512.01295}, 2025.

\bibitem{li2025aac}
X.~Li, D.~Huang, J.~Li, H.~Cai, Z.~Zhou, W.~Dong, X.~Wang, and Y.~Liu, ``A
  vision for access control in llm-based agent systems,'' \emph{arXiv preprint
  arXiv:2510.11108}, 2025.

\bibitem{owasp2025llmtop10}
{OWASP Foundation}, ``{OWASP} top 10 for large language model applications,''
  \url{https://owasp.org/www-project-top-10-for-large-language-model-applications/},
  2025.

\bibitem{zdi2026codingagents}
{Zero Day Initiative}, ``{Pwn2Own Berlin} 2026: Coding agents category rules
  and day one results,''
  \url{https://www.thezdi.com/blog/2026/5/13/pwn2own-berlin-2026-day-one-results},
  2026, category and eligibility rules per the event announcement; targets
  Claude Code, OpenAI Codex, and Cursor.

\bibitem{kovacs2026pwn2own}
E.~Kovacs, ``Hackers earn over \$1.29 million at {Pwn2Own Berlin} 2026,''
  SecurityWeek, \url{https://www.securityweek.com/}, 2026, 47 unique zero-days;
  total payout \$1{,}298{,}250.

\bibitem{zhong2026yolofs}
S.~W. Zhong, J.~Liao, J.~Liu, M.~Zheng, A.~C. Arpaci-Dusseau, and R.~H.
  Arpaci-Dusseau, ``Don't let ai agents yolo your files: Shifting information
  and control to filesystems for agent safety and autonomy,'' \emph{arXiv
  preprint arXiv:2604.13536}, 2026.

\bibitem{wu2025isolategpt}
Y.~Wu, F.~Roesner, T.~Kohno, N.~Zhang, and U.~Iqbal, ``Isolategpt: An execution
  isolation architecture for llm-based agentic systems,'' in \emph{Network and
  Distributed System Security Symposium (NDSS)}, 2025, arXiv:2403.04960.

\bibitem{meng2025cellmate}
L.~Meng, H.~Feng, I.~Shumailov, and E.~Fernandes, ``cellmate: Sandboxing
  browser ai agents,'' \emph{arXiv preprint arXiv:2512.12594}, 2025.

\bibitem{piao2025agentbay}
Y.~Piao, H.~Min, H.~Su, L.~Zhang, L.~Wang, Y.~Yin, X.~Wu, Z.~Xu, L.~Qu, H.~Li
  \emph{et~al.}, ``Agentbay: A hybrid interaction sandbox for seamless human-ai
  intervention in agentic systems,'' \emph{arXiv preprint arXiv:2512.04367},
  2025.

\bibitem{dong2026deltabox}
Y.~Dong, J.~He, S.~Liu, Y.~Hou, D.~Du, Z.~Xu, S.~Yu, B.~Yang, Y.~Xia, and
  H.~Chen, ``Deltabox: Scaling stateful ai agents with millisecond-level
  sandbox checkpoint/rollback,'' \emph{arXiv preprint arXiv:2605.22781}, 2026.

\bibitem{yan2025faulttolerant}
B.~Yan, ``Fault-tolerant sandboxing for ai coding agents: A transactional
  approach to safe autonomous execution,'' \emph{arXiv preprint
  arXiv:2512.12806}, 2025.

\bibitem{marchand2026sandboxescape}
R.~Marchand, A.~O. Cathain, J.~Wynne, P.~M. Giavridis, S.~Deverett,
  J.~Wilkinson, J.~Gwartz, and H.~Coppock, ``Quantifying frontier llm
  capabilities for container sandbox escape,'' \emph{arXiv preprint
  arXiv:2603.02277}, 2026.

\bibitem{guo2024redcode}
C.~Guo, X.~Liu, C.~Xie, A.~Zhou, Y.~Zeng, Z.~Lin, D.~Song, and B.~Li,
  ``Redcode: Risky code execution and generation benchmark for code agents,''
  in \emph{Advances in Neural Information Processing Systems (NeurIPS) Datasets
  and Benchmarks Track}, 2024, arXiv:2411.07781.

\bibitem{rabin2025sandboxeval}
R.~Rabin, J.~Hostetler, S.~McGregor, B.~Weir, and N.~Judd, ``Sandboxeval:
  Towards securing test environment for untrusted code,'' \emph{arXiv preprint
  arXiv:2504.00018}, 2025.

\bibitem{ruan2023toolemu}
Y.~Ruan, H.~Dong, A.~Wang, S.~Pitis, Y.~Zhou, J.~Ba, Y.~Dubois, C.~J. Maddison,
  and T.~Hashimoto, ``Identifying the risks of {LM} agents with an
  {LM}-emulated sandbox,'' in \emph{International Conference on Learning
  Representations (ICLR)}, 2024, arXiv:2309.15817.

\bibitem{zhan2024injecagent}
Q.~Zhan, Z.~Liang, Z.~Ying, and D.~Kang, ``{InjecAgent}: Benchmarking indirect
  prompt injections in tool-integrated large language model agents,'' in
  \emph{Findings of the Association for Computational Linguistics: ACL}, 2024,
  arXiv:2403.02691.

\bibitem{debenedetti2024agentdojo}
E.~Debenedetti, J.~Zhang, M.~Balunovi{\'c}, L.~Beurer-Kellner, M.~Fischer, and
  F.~Tram{\`e}r, ``{AgentDojo}: A dynamic environment to evaluate prompt
  injection attacks and defenses for {LLM} agents,'' in \emph{Neural
  Information Processing Systems Datasets and Benchmarks Track (NeurIPS D\&B)},
  2024, arXiv:2406.13352.

\bibitem{buhler2025agentbound}
C.~Buhler, M.~Biagiola, L.~D. Grazia, and G.~Salvaneschi, ``Agentbound:
  Securing execution boundaries of ai agents,'' \emph{arXiv preprint
  arXiv:2510.21236}, 2025.

\bibitem{santosgrueiro2026portico}
I.~Santos-Grueiro, ``Lingering authority: Revocable resource-and-effect
  capabilities for coding agents,'' \emph{arXiv preprint arXiv:2606.22504},
  2026.

\bibitem{ji2026seagent}
Z.~Ji, D.~Wu, W.~Jiang, P.~Ma, Z.~Li, Y.~Gao, S.~Wang, and Y.~Li, ``Taming
  various privilege escalation in llm-based agent systems: A mandatory access
  control framework,'' \emph{arXiv preprint arXiv:2601.11893}, 2026.

\bibitem{uchibeke2026oap}
U.~Uchibeke, ``Before the tool call: Deterministic pre-action authorization for
  autonomous ai agents,'' \emph{arXiv preprint arXiv:2603.20953}, 2026.

\bibitem{palumbo2026forge}
N.~Palumbo, S.~Choudhary, J.~Choi, G.~Amir, P.~Chalasani, and S.~Jha, ``Formal
  policy enforcement for real-world agentic systems,'' \emph{arXiv preprint
  arXiv:2602.16708}, 2026.

\bibitem{chen2026shellsieve}
C.~Chen and Z.~Lin, ``One goal, many commands: Characterizing denylist
  fragility in ai agents,'' \emph{arXiv preprint arXiv:2606.15549}, 2026.

\bibitem{lilienthal2025toctou}
D.~Lilienthal and S.~Hong, ``Mind the gap: Time-of-check to time-of-use
  vulnerabilities in llm-enabled agents,'' \emph{arXiv preprint
  arXiv:2508.17155}, 2025.

\bibitem{jiang2026browsertoctou}
L.~Jiang, Z.~Liu, H.~Luo, and Z.~Lin, ``Atomicity for agents: Exposing,
  exploiting, and mitigating toctou vulnerabilities in browser-use agents,''
  \emph{arXiv preprint arXiv:2603.00476}, 2026.

\bibitem{huang2026mcpthreat}
C.~Huang, X.~Huang, N.~P. Tran, and A.~M. Fard, ``Model context protocol threat
  modeling and analyzing vulnerabilities to prompt injection with tool
  poisoning,'' \emph{arXiv preprint arXiv:2603.22489}, 2026.

\bibitem{hou2025mcplandscape}
X.~Hou, Y.~Zhao, S.~Wang, and H.~Wang, ``Model context protocol (mcp):
  Landscape, security threats, and future research directions,'' \emph{arXiv
  preprint arXiv:2503.23278}, 2025.

\bibitem{hasan2025mcpfirstglance}
M.~M. Hasan, H.~Li, E.~Fallahzadeh, G.~K. Rajbahadur, B.~Adams, and A.~E.
  Hassan, ``Model context protocol (mcp) at first glance: Studying the security
  and maintainability of mcp servers,'' \emph{arXiv preprint arXiv:2506.13538},
  2025.

\bibitem{xing2026mcpguard}
W.~Xing, Z.~Qi, Y.~Qin, Y.~Li, C.~Chang, J.~Yu, C.~Lin, Z.~Xie, and M.~Han,
  ``Mcp-guard: A multi-stage defense-in-depth framework for securing model
  context protocol in agentic ai,'' \emph{arXiv preprint arXiv:2508.10991},
  2026.

\bibitem{wang2026hciframing}
P.~Wang, Y.~Li, and Y.~Tian, ``Reframing llm agent security as an agent-human
  interaction problem,'' \emph{arXiv preprint arXiv:2605.24309}, 2026.

\bibitem{mayoralvilches2026csi}
V.~Mayoral-Vilches, F.~Balassone, M.~Sanz-Gomez, P.~Z. Landa, D.~S. Prieto,
  M.~O. Alvarez, D.~Quarta, and M.~Pinzger, ``Towards cybersecurity
  superintelligence (csi): What's the best harness for cybersecurity?''
  \emph{arXiv preprint arXiv:2605.28334}, 2026.

\bibitem{south2025delegation}
T.~South, S.~Marro, T.~Hardjono, R.~Mahari, C.~D. Whitney, D.~Greenwood,
  A.~Chan, and A.~Pentland, ``Authenticated delegation and authorized ai
  agents,'' \emph{arXiv preprint arXiv:2501.09674}, 2025.

\bibitem{zhou2026capabilitybinding}
Z.~Zhou, ``Governing dynamic capabilities: Cryptographic binding and
  reproducibility verification for ai agent tool use,'' \emph{arXiv preprint
  arXiv:2603.14332}, 2026.

\bibitem{sequeira2026agentsentry}
R.~Sequeira, S.~Damianakis, U.~Iqbal, and K.~Psounis, ``Agent-sentry: Bounding
  llm agents via execution provenance,'' \emph{arXiv preprint
  arXiv:2603.22868}, 2026.

\bibitem{abdelnabi2025firewalls}
S.~Abdelnabi, A.~Gomaa, E.~Bagdasarian, P.~O. Kristensson, and R.~Shokri,
  ``Firewalls to secure dynamic llm agentic networks,'' \emph{arXiv preprint
  arXiv:2502.01822}, 2025.

\bibitem{chen2025securevibebench}
J.~Chen, H.~Huang, Y.~Lyu, J.~An, J.~Shi, C.~Yang, T.~Zhang, H.~Tian, Y.~Li,
  Z.~Li, X.~Zhou, X.~Hu, and D.~Lo, ``Securevibebench: Benchmarking secure vibe
  coding of ai agents via reconstructing vulnerability-introducing scenarios,''
  \emph{arXiv preprint arXiv:2509.22097}, 2025.

\bibitem{qu2026overeager}
Y.~Qu, Y.~Zhang, Y.~Zhang, G.~Deng, Y.~Li, L.~Y. Zhang, and Y.~Liu, ``Overeager
  coding agents: Measuring out-of-scope actions on benign tasks,'' \emph{arXiv
  preprint arXiv:2605.18583}, 2026.

\bibitem{li2026secureskills}
Z.~Li, J.~Wu, X.~Ling, X.~Cui, and T.~Luo, ``Towards secure agent skills:
  Architecture, threat taxonomy, and security analysis,'' \emph{arXiv preprint
  arXiv:2604.02837}, 2026.

\bibitem{greshake2023indirect}
K.~Greshake, S.~Abdelnabi, S.~Mishra, C.~Endres, T.~Holz, and M.~Fritz, ``Not
  what you've signed up for: Compromising real-world {LLM}-integrated
  applications with indirect prompt injection,'' in \emph{ACM Workshop on
  Artificial Intelligence and Security (AISec)}, 2023, arXiv:2302.12173.

\bibitem{liu2023promptinjapps}
Y.~Liu, G.~Deng, Y.~Li, K.~Wang, T.~Zhang, Y.~Liu, H.~Wang, Y.~Zheng, and
  Y.~Liu, ``Prompt injection attack against {LLM}-integrated applications,''
  \emph{arXiv preprint arXiv:2306.05499}, 2023.

\bibitem{zou2023universal}
A.~Zou, Z.~Wang, N.~Carlini, M.~Nasr, J.~Z. Kolter, and M.~Fredrikson,
  ``Universal and transferable adversarial attacks on aligned language
  models,'' \emph{arXiv preprint arXiv:2307.15043}, 2023.

\bibitem{yuan2024rjudge}
T.~Yuan, Z.~He, L.~Dong, Y.~Wang, R.~Zhao, T.~Xia, L.~Xu, B.~Zhou, F.~Li,
  Z.~Zhang, R.~Wang, and G.~Liu, ``{R-Judge}: Benchmarking safety risk
  awareness for {LLM} agents,'' in \emph{Findings of the Association for
  Computational Linguistics: EMNLP}, 2024, arXiv:2401.10019.

\bibitem{zhu2025melon}
K.~Zhu, X.~Yang, J.~Wang, W.~Guo, and W.~Y. Wang, ``{MELON}: Provable defense
  against indirect prompt injection attacks in {AI} agents,'' \emph{arXiv
  preprint arXiv:2502.05174}, 2025.

\bibitem{shi2025promptarmor}
T.~Shi, K.~Zhu, Z.~Wang, Y.~Jia, W.~Cai, W.~Liang, H.~Wang, H.~Alzahrani,
  J.~Lu, K.~Kawaguchi, B.~Alomair, X.~Zhao, W.~Y. Wang, N.~Gong, W.~Guo, and
  D.~Song, ``{PromptArmor}: Simple yet effective prompt injection defenses,''
  \emph{arXiv preprint arXiv:2507.15219}, 2025.

\bibitem{weng2026argus}
S.~Weng, Y.~Feng, J.~Zhang, X.~Xie, J.~Yu, and J.~Liu, ``{ARGUS}: Defending
  {LLM} agents against context-aware prompt injection,'' \emph{arXiv preprint
  arXiv:2605.03378}, 2026.

\bibitem{abdelnabi2026alwaysfall}
S.~Abdelnabi and E.~Bagdasarian, ``{AI} agents may always fall for prompt
  injections,'' \emph{arXiv preprint arXiv:2605.17634}, 2026.

\end{thebibliography}
